\documentclass[referee]{aa} 
\usepackage{graphicx}
\usepackage{graphicx}
\usepackage{txfonts}
\usepackage{subfigure}
\usepackage{natbib}
\bibpunct{(}{)}{;}{a}{}{,} 

\begin{document}
\title{Pulsating low-mass white dwarfs in the frame of new evolutionary 
sequences} 
\subtitle{IV. The secular rate of period change}

\author{Leila M. Calcaferro\inst{1,2},
        Alejandro H. C\'orsico\inst{1,2}, \and   
        Leandro G. Althaus\inst{1,2}}
\institute{$^{1}$ Grupo de Evoluci\'on Estelar y Pulsaciones,
  Facultad de Ciencias Astron\'omicas y Geof\'isicas,
  Universidad Nacional de La Plata, Paseo del Bosque s/n,
  1900, La Plata, Argentina\\
  $^{2}$ Instituto de Astrof\'isica La Plata,
  CONICET-UNLP, Paseo del Bosque s/n, 1900, La Plata, Argentina\\
           \email{lcalcaferro,acorsico,althaus@fcaglp.unlp.edu.ar}     
           }
\date{Received ; accepted }

\abstract{An increasing number of low-mass ($M_{\star}/M_{\sun}
  \lesssim 0.45$) and extremely low-mass (ELM, $M_{\star}/M_{\sun}
  \lesssim 0.18-0.20$) white-dwarf stars are currently discovered in
  the  field of the Milky  Way. Some of these stars exhibit
  long-period $g$-mode pulsations, and are called ELMV variable
  stars. Also, some low-mass pre-white dwarf stars
  show short-period $p$-mode (and likely radial-mode) photometric
  variations, and are designated as pre-ELMV variable stars. The
  existence of these new classes of pulsating white dwarfs and
  pre-white dwarfs opens the prospect of exploring the binary
  formation channels of these low-mass white dwarfs through
  asteroseismology.}{We present  a theoretical assessment of the
  expected temporal rates of change of periods ($\dot{\Pi}$) for such stars,
  based on fully evolutionary low-mass He-core white
  dwarf and pre-white dwarf models.}{Our analysis is based on a large set
  of adiabatic periods of radial and nonradial pulsation modes
  computed on a suite of low-mass He-core white dwarf and pre-white
  dwarf models with masses  ranging  from $0.1554$   to  $0.4352
  M_{\sun}$, which were derived by computing  the non-conservative
  evolution of a binary system consisting of an  initially $1
  M_{\sun}$ ZAMS star and a $1.4 M_{\sun}$ neutron star companion.}{We
  compute the secular rates of period change of radial ($\ell= 0$) and
  nonradial ($\ell= 1, 2$) $g$ and $p$ modes for stellar models
  representative of ELMV and pre-ELMV stars, as well as for stellar
  objects that are evolving just before the occurrence of CNO flashes
  at the early cooling branches. We found that the theoretically
  expected magnitude of $\dot{\Pi}$ of $g$ modes for pre-ELMVs are by
  far larger than for ELMVs. In turn, $\dot{\Pi}$ of $g$ modes for
  models evolving before the occurrence of CNO flashes are larger than
  the maximum values of the rates of period change predicted for
  pre-ELMV stars. Regarding $p$ and radial modes, we found that the
  larger absolute values of $\dot{\Pi}$ correspond to pre-ELMV
  models.}{We conclude that any eventual measurement of a rate
  of period change for a given pulsating low-mass pre-white dwarf or
  white dwarf star could shed light about its evolutionary status.
  Also, in view of the systematic difficulties in the spectroscopic
  classification of stars of the ELM Survey, an eventual
  measurement of $\dot{\Pi}$ could help to confirm that a given
  pulsating star is an authentic low-mass white dwarf and not a star from
  another stellar population.}

\titlerunning{The rates of period change of low-mass white dwarfs}
   \maketitle
%

\section{Introduction}
\label{introduction}

Low-mass  ($M_{\star}/M_{\odot}  \lesssim  0.45$) white dwarfs (WD)
are probably produced  by strong  mass-loss episodes  at the red giant
branch phase of low-mass  stars in  binary systems before the He-flash
onset \citep{review}. Since He burning is avoided, they are expected
to harbor  He cores,  in  contrast to  average-mass ($M_{\star} \sim
0.6 M_{\odot}$) C/O-core  WDs. In   particular, binary evolution is
the most likely origin for  the so-called  extremely low-mass (ELM)
WDs, which  have masses $M_{\star} \sim  0.18-0.20 M_{\odot}$.  The
evolution of low-mass  WDs is strongly dependent on their stellar mass
and the occurrence of element diffusion processes
\citep{2001MNRAS.323..471A}.
\citet{2001MNRAS.323..471A,2013A&A...557A..19A,2016A&A...595A..35I}
have found that  element diffusion leads to a dichotomy regarding the
thickness of the H envelope,  which translates into a dichotomy in the
age   of  low-mass  He-core  WDs. Specifically,  for   stars  with
$M_{\star}  \gtrsim 0.18 -  0.20 M_{\odot}$,  the WD  progenitor
experiences multiple diffusion-induced CNO  thermonuclear flashes that
consume most of the H content of the  envelope, and as a result, the
remnant enters its final cooling track with a  very thin H
envelope. The resulting object is unable to sustain substantial
nuclear burning while it cools,  and its evolutionary timescale is rather
short   ($\sim 10^7-10^8$ yr). On the contrary,  if $M_{\star}
\lesssim 0.18 - 0.20M_{\odot}$, the  WD progenitor does not experience
H  flashes at all, and the  remnant enters its terminal cooling branch
with  a thick  H envelope.  In this case,  residual H  nuclear burning
via  $pp$-chain becomes the  main  energy source, that ultimately
slows down  the evolution of the star,  in which  case the cooling
timescale is  of the order of $\sim 10^9$  yrs.  The age dichotomy has
been also suggested by  observations of those  low-mass He-core WDs
that are companions  to millisecond pulsars
\citep{2003A&A...403.1067B}. 

In the past few years, numerous low-mass WDs, including  ELM  WDs,
have been detected through the ELM survey and the SPY and WASP
surveys \citep[see][]{2009A&A...505..441K,  2010ApJ...723.1072B,
  2012ApJ...744..142B, 2011MNRAS.418.1156M, 2011ApJ...727....3K,
  2012ApJ...751..141K, 2013ApJ...769...66B, 2014ApJ...794...35G,
  2015MNRAS.446L..26K, 2015ApJ...812..167G, 2016ApJ...818..155B,
  2016ApJ...824...46B}.  The interest in low-mass WDs has been greatly
promoted by the discovery of pulsations in some of them
\citep{2012ApJ...750L..28H, 2013ApJ...765..102H,2013MNRAS.436.3573H,
  2015MNRAS.446L..26K, 2015ASPC..493..217B,
  2016arXiv161206390B}\footnote{The star SDSS J135512+195645
  discovered by \citet{2016arXiv161206390B} is likely a high-amplitude
  $\delta$ Scuti pulsator with an overestimated surface gravity,  as
    pointed by these authors.}.
The discovery of
pulsating  low-mass WDs (hereinafter ELMVs\footnote{For simplicity,
  here  and  throughout the  paper  we  refer  to the  pulsating
  low-mass  WDs  as  ELMVs, even  if $M_{\star} \gtrsim 0.18-0.20
  M_{\sun}$.})  constitutes a unique chance for probing the interior
of  these stars and eventually to test their formation scenarios by
employing the methods of asteroseismology. Theoretical  adiabatic
pulsational analyzes of these stars \citep{2010ApJ...718..441S,
  2012A&A...547A..96C, 2014A&A...569A.106C} show that $g$ modes  in
ELM WDs are restricted mainly to  the core regions,  providing the
chance to constrain  the core  chemical structure.   Also,
nonadiabatic  stability computations \citep{2012A&A...547A..96C,
  2013ApJ...762...57V,2016A&A...585A...1C} show that many unstable $g$
and $p$  modes are excited by  a combination of the $\kappa-\gamma$
mechanism \citep{1989nos..book.....U} and the ``convective driving''
mechanism \citep{1991MNRAS.251..673B},  both of them acting at the
H-ionization  zone. In addition,     the $\varepsilon$ mechanism due
to stable H burning could contribute to destabilize some short-period
$g$    modes  in ELM WDs \citep{2014ApJ...793L..17C}.

Apart from ELMVs, pulsations in  several objects that are likely the
precursors  of low-mass WD stars have been detected in the last years
\citep{2013Natur.498..463M, 2014MNRAS.444..208M,2016ApJ...821L..32Z,
  2016ApJ...822L..27G,2016A&A...587L...5C}\footnote{The nature of the
  variable stars reported by \citet{2016A&A...587L...5C} is unclear,
  as they  could be precursors
  of low-mass WD stars or, alternatively,  $\delta$ Scuti/SX Phe-like
  stars,  as pointed by these authors.}. Nonadiabatic stability
computations for radial modes
\citep{2013MNRAS.435..885J} and nonradial  $p$ and  $g$ modes
\citep{2016A&A...588A..74C, 2016ApJ...822L..27G,2016A&A...595L..12I}
have revealed that the excitation of pulsations in these pre-WDs is
the $\kappa-\gamma$ mechanism acting mainly at the zone of the second
partial ionization of He, with a weaker contribution from the region
of the first partial ionization  of He and the partial  ionization of
H. So, the abundance of He in the envelopes of this new class of
pulsating stars (hereinafter pre-ELMVs\footnote{Here and throughout
  the  paper  we  refer  to the  pulsating low-mass  pre-WDs  as
  pre-ELMVs, even  if $M_{\star} \gtrsim 0.18-0.20 M_{\sun}$.})  is a
crucial ingredient for destabilizing the pulsation modes
\citep{2016A&A...588A..74C,2016A&A...595L..12I}.

The  $g$-mode pulsation periods ($\Pi$) of WD stars experience a
secular drift  as they  cool,  giving place to a detectable  rate  of
period  change, $\dot{\Pi}\equiv d\Pi/dt$. Specifically, as  the
temperature in the core of a WD decreases,  the plasma increases  its
degree of degeneracy so that the  Brunt-V\"ais\"al\"a  (buoyancy)
frequency  ---  the  critical frequency  of   $g$-mode  pulsations
\citep{1989nos..book.....U}  --- decreases,  and the  pulsational
spectrum  of the  star is  shifted to longer periods.  On the other
hand, gravitational contraction (if  present) acts  in  the opposite
direction,  favoring the  shortening of the pulsation periods.  The
competition  between the  increasing degeneracy and gravitational
contraction gives  rise to a detectable $\dot{\Pi}$.  In particular,
it has been  shown by \cite{Wingetet83} that the  rate of change of
the $g$-mode  pulsation periods is related to the  rate of  change of
the temperature at the region  of the period formation,  $\dot{T}$,
and the rate of change of the  stellar radius,  $\dot{R_{\star}}$,
according   to the following order-of-magnitude expression:

\begin{equation}
  \frac{\dot{\Pi}}{\Pi} \approx -a \frac{\dot{T}}{T} +
  b \frac{\dot{R_{\star}}}{R_{\star}},
\label{eq-dotp}
\end{equation}

\noindent where $a$  and $b$ are constants whose values  depend on the
details of the WD modeling  (however $a, b \approx 1$).  The first
term in Eq.~(\ref{eq-dotp}) corresponds to the rate of change in
period induced by the  cooling of the WD, and since $\dot{T}<0$, it is
a positive contribution. The second term represents the rate of change
due to gravitational contraction ($\dot{R_{\star}} < 0$),  and it is a
negative contribution. 

In the cases in which robust measurements of secular period drifts
  of pulsating WDs
  can be achieved, a number of important applications can be
  ---in principle--- carried out
\citep{2003ApJ...594..961M}. In particular, the derived
values of  $\dot{\Pi}$ could help in calibrating the WD cooling
curves, thus
reducing  the theoretical uncertainties of WD cosmochronology to
constrain  the age of the Galactic disk
\citep[e.g.,][]{2006AJ....131..571H},  halo
\citep[e.g.,][]{1998ApJ...503..239I}, and Galactic globular clusters
\citep[e.g.,][]{2013Natur.500...51H} and open clusters
\citep[][]{2010Natur.465..194G}. The measurement of $\dot{\Pi}$  also
could allow us to infer the chemical composition of the core of a WD
\citep{2005ApJ...634.1311K}.  This is because the rate of cooling of
WDs, and so, the rate of  period change of a given pulsation mode,
depend primarily on the core composition and the stellar mass.  At fixed
mass, $\dot{\Pi}$ is larger for higher mean atomic weight  of the
core. This allows to place constraints on the core chemical
composition. Another possible application of the measurement of
$\dot{\Pi}$ is the detection of planets. The orbital motion of a
pulsating WD around the center of mass of the system due to  the
possible presence of a planet modify the light travel time of the
pulses.  As a result, the observed arrival time on Earth changes,
thus providing an alternative  method to detect the
planet \citep{2008ApJ...676..573M}.
Finally, the rates of  period change  in WDs allow, in principle, to
place constraints on axions \citep{Isern92,2001NewA....6..197C,
  2008ApJ...675.1512B,2012MNRAS.424.2792C,2012JCAP...12..010C,2016JCAP...07..036C, 2016JCAP...08..062B},
neutrinos   \citep{2004ApJ...602L.109W,2014JCAP...08..054C},  and  the
possible  secular  rate of  variation  of  the gravitational  constant
\citep{2013JCAP...06..032C}. Note that, however, in order to
  establish robust
constraints based on the rate of change of periods, it is necessary
to know with some degree of accuracy the total mass, the effective
temperature, the core composition, and the envelope layering of the target
star \citep{FB08}.

The rate of change of the periods can  be measured, in principle,  by
monitoring a  pulsating WD  over a long time interval when one or more
very stable pulsation periods are present  in their  power spectrum.
In the  case of  pulsating DA   (H-rich atmosphere) and DB (He-rich
atmosphere) WDs, also called  DAV and DBV stars, respectively, cooling
dominates over  gravitational  contraction, in such a way that the
second term in Eq.~(\ref{eq-dotp}) is usually negligible, and only
positive values of the  observed rate of change of period are expected
\citep{WK08,FB08,review}. For C/O-core DAVs, the expected rates of
period  change are in the range $10^{-15}-10^{-16}$ s/s
\citep{1992ApJ...391L..33B,B96}, in excellent agreement with the
measured values for G117$-$B15A \citep[$\dot{\Pi}= 4.19 \pm 0.73
  \times 10^{-15}$ s/s,][]{Kea12},  R548 \citep[$\dot{\Pi}= 3.3 \pm
  1.1 \times 10^{-15}$ s/s,] []{2013ApJ...771...17M} and L19$-$2
\citep[$\dot{\Pi}= 3.0 \pm 0.6 \times 10^{-15}$
  s/s,][]{2015ASPC..493..199S}, although in strong conflict with the
value derived for WD 0111+0018 \citep[$\dot{\Pi} > 10^{-12}$
  s/s,][]{2013ApJ...766...42H}.  For DBVs, an estimate of the rate of
period change has been obtained for PG 1351+489 \citep[$\dot{\Pi}= 2.0
  \pm 0.9 \times 10^{-13}$ s/s,][]{2011MNRAS.415.1220R},  in line with
the theoretical expectations  \citep[$\dot{\Pi} \sim
  10^{-13}-10^{-14}$ s/s;][]{2004ApJ...602L.109W,2004A&A...428..159C}.
In the case of pulsating hot WD and pre-WD stars, also called GW Vir
or pulsating PG1159 stars (He-, C-, and O-rich atmosphere),
theoretical models predict rates of period change in the range
$10^{-11}-10^{-12}$ s/s
\citep{1994ApJ...427..415K,2006A&A...454..863C,
  2008A&A...478..869C}. For the high effective temperatures
characterizing the GW Vir instability strip, gravitational contraction
is still  significant, to such a degree that its influence on
$\dot{\Pi}$ can overcome the effects of cooling. In this case the
second term in  Eq. (\ref{eq-dotp}) is not negligible and, therefore,
either positive or negative values of $\dot{\Pi}$ are possible.
$\dot{\Pi}$   for several $g$ modes has been measured in the case of
the  prototypical GW Vir star, PG1159$-$035
\citep{2008A&A...489.1225C}. The star exhibits a mixture of positive
and negative $\dot{\Pi}$ values  of large magnitude, up to $\sim 4
\times 10^{-10}$ s/s.   In particular, the rate of period change of
the mode with period $\Pi= 517.1$ s is $\dot{\Pi}= 1.52 \pm 0.05\times
10^{-10}$ s/s,  an order of magnitude larger than the theoretical
predictions
\citep{1994ApJ...427..415K,2006A&A...454..863C,2008A&A...478..869C}.
\citet{2008ApJ...677L..35A} have found that this discrepancy could be
alleviated if PG1159$-$035 is  characterized by a thin He-rich
envelope, leading to remarkably large  magnitudes of the rates of
period change.  A measurement of $\dot{\Pi}$ in another GW Vir star,
PG0112+200,  has been carried out by \citet{2011A&A...528A...5V}.  The
derived rates of period change are much larger than  those predicted
by theoretical models \citep{2007A&A...475..619C},  calling for the
presence of other mechanism(s) apart from neutrino cooling to  explain
the disagreement. In particular, a mechanism that could be  playing a
dominant role is resonant mode coupling induced by the rotation
\citep{2011A&A...528A...5V}.  A cautionary note regarding the
  interpretation of the measured rates of period change in pulsating WDs is
  needed here. The studies by \citet{2013ApJ...766...42H} for the
  DAV star WD 0111+0018, and \citet{2011A&A...528A...5V} for the GW Vir
  star PG0112+200 (among others), indicate that our understanding
  of the rates of period change in pulsating WDs is far from
  complete, and this should be kept in mind when using $\dot{\Pi}$
  in the applications mentioned before.

In this paper,  the fourth one of a series devoted to low-mass WD and
pre-WD stars, we present for the first time a detailed assessment of
the theoretical temporal rates of period change of ELMV and pre-ELMV
stars. According to the theoretically estimated rates of cooling of
these stars \citep{2013A&A...557A..19A}, low-mass WDs cool slower than
low-mass pre-WDs. On these grounds, it is expected that ELMVs will
have smaller rates of period change than pre-ELMVs. Therefore, the
eventual measurement of the rate of period change for a given
pulsating star could be potentially useful to distinguish in which
evolutionary stage the star is.  Also, an eventual measurement of
$\dot{\Pi}$ could help, in principle, to distinguish genuine ELM WDs
$(M_{\star} \lesssim 0.18-0.20 M_{\odot})$ that have thick H envelopes
and long cooling timescales, from low-mass WDs $(M_{\star} \gtrsim
0.18-0.20 M_{\odot})$, characterized by thinner H envelopes and
shorter cooling timescales.  However, we must keep in mind that the
cooling rates of this kind of stars could be so slow, that any secular
period change would be very difficult to detect. Notwithstanding, some
of these stars may not be on their terminal cooling tracks but rather
may be on the pre-WD stage, or even going to a CNO flash, and thus
have much higher (and much more easily measurable) $\dot{\Pi}$ values.
In other words, the detection of any significant secular period change
would be strong evidence for the object to be not on its final cooling
track as an ELMV star.  Although the measurement of $\dot{\Pi}$  for
any of these stars is not expected shortly, it could be achieved in
the next years by means of continuous photometric monitoring of these
objects.

The paper is organized as  follows.  In Sect.  \ref{modelling} we
briefly describe  our  numerical  tools   and  the  main  ingredients
of  the evolutionary sequences  we employ to  assess the rates of
period change of low-mass He-core WDs and pre-WDs.   In Sect.
\ref{rates} we present in detail our results of $\dot{\Pi}$ for ELMV
and pre-ELMV models. In particular, we study the dependence of the
rates of period change of nonradial dipole ($\ell= 1$) $g$ and $p$
modes, and radial ($\ell= 0$) modes with the stellar mass and the
effective temperature. We expand the analysis by including the
assessment of $\dot{\Pi}$ for stellar models evolving at stages
previous of the development of thermonuclear CNO flashes during the
early-cooling phase.  Finally,    in Sect. \ref{conclusions} we
summarize  the main findings of the paper.

\section{Modeling}
\label{modelling}

\subsection{Evolutionary code}
\label{evol-code}

The fully evolutionary models of low-mass He-core WD and pre-WD stars
on which this work is based were generated with  the {\tt LPCODE}
stellar evolution code. {\tt LPCODE} computes in detail the complete
evolutionary stages leading to  WD formation, allowing one to study
the WD and pre-WD evolution in a consistent way with the expectations
of the evolutionary  history of progenitors.  Details of {\tt LPCODE}
can be found in
\citet{2005A&A...435..631A,2009A&A...502..207A,2013A&A...557A..19A,
  2015A&A...576A...9A,2016arXiv161106191A} and references
therein. Here, we mention  only  those physical ingredients which  are
relevant for  our  analysis  of  low-mass, He-core WD and pre-WD stars
\citep[see][for details]{2013A&A...557A..19A}.  The standard  Mixing
Length  Theory (MLT)   for convection   in the version ML2 is used
\citep{1990ApJS...72..335T}. The metallicity of the progenitor stars
has been  assumed to be $Z = 0.01$.  Radiative opacities for arbitrary
metallicity in the range from 0 to  0.1 are from the OPAL project
\citep{1996ApJ...464..943I}.  Conductive opacities are those of
\citet{2007ApJ...661.1094C}. The  equation of state during the main
sequence evolution  is that of OPAL for H- and He-rich compositions.
Neutrino  emission   rates   for   pair, photo,   and bremsstrahlung
processes have been taken from \citet{1996ApJS..102..411I},   and for
plasma processes we included the treatment of
\citet{1994ApJ...425..222H}. For the WD  regime we have employed  an
updated version of the \citet{1979A&A....72..134M} equation of
state. The nuclear network   takes into account 16 elements and 34
thermonuclear  reaction  rates   for  pp-chains,  CNO bi-cycle,  He
burning, and C ignition. Time-dependent element diffusion due to
gravitational  settling and chemical  and thermal diffusion  of
nuclear  species  has been  taken into account following  the
multicomponent  gas  treatment  of
\citet{1969fecg.book.....B}. Abundance changes  have been computed
according to element diffusion, nuclear reactions,  and convective
mixing.  This detailed treatment  of abundance  changes by  different
processes during the WD regime constitutes a key aspect in the
evaluation of the importance of  residual nuclear burning for the
cooling of low-mass WDs.

\begin{figure} 
\begin{center}
\includegraphics[clip,width=9 cm]{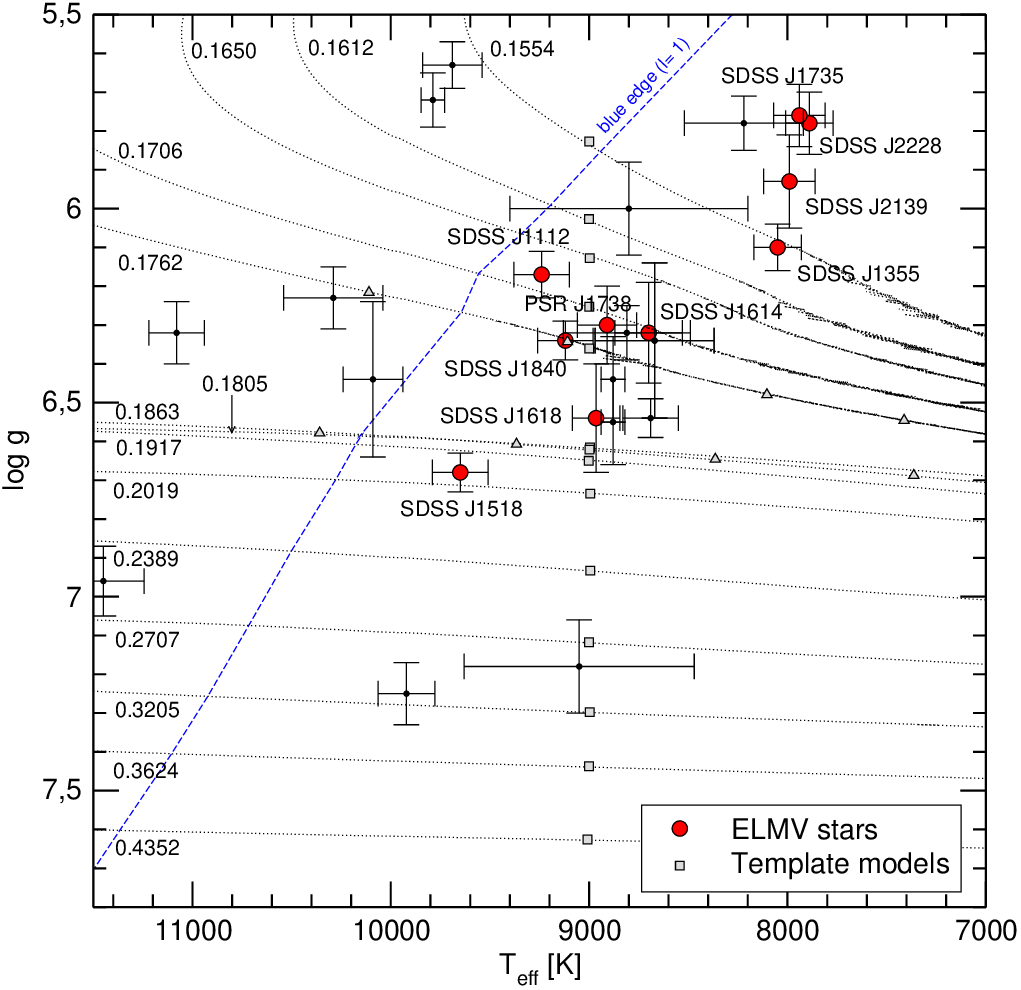} 
\caption{$T_{\rm eff} - \log g$ plane showing the low-mass He-core WD
  evolutionary tracks (final cooling branches) of
  \citet{2013A&A...557A..19A}.  Numbers correspond to the stellar mass
  of each sequence.  The location of the ten known ELMVs
  \citep{2012ApJ...750L..28H,
    2013ApJ...765..102H,2013MNRAS.436.3573H,2015MNRAS.446L..26K,2015ASPC..493..217B,2016arXiv161206390B}
  are marked with large red circles ($T_{\rm eff}$ and $\log g$
  computed with 3D model atmospheres corrections).  Stars observed not
  to vary
  \citep{2012PASP..124....1S,2012ApJ...750L..28H,2013ApJ...765..102H,2013MNRAS.436.3573H}
  are depicted with small black circles. The Gray squares and
  triangles on the evolutionary tracks indicate the location of the
  template models analyzed in the text.  The dashed blue line
  corresponds to the blue edge of the instability domain of $\ell= 1$
  $g$ modes according to the nonadiabatic computations of
  \citet{2016A&A...585A...1C} using ML2 ($\alpha= 1.0$) version of the
  MLT theory of convection.}
\label{HR-ELMs} 
\end{center}
\end{figure}

\subsection{Pulsation code}
\label{pulsa-code}

The rates of period change of  radial modes and nonradial
$p$ and $g$ modes computed in this work
were derived from the large set of adiabatic pulsation periods
presented in \citet{2014A&A...569A.106C}. These periods were computed
employing the adiabatic radial and nonradial versions of the
{\tt LP-PUL} pulsation code described in detail 
in \citet{2006A&A...454..863C,2014A&A...569A.106C}, which is
coupled to the {\tt LPCODE} evolutionary  code. The {\tt LP-PUL} pulsation
code  is  based  on  a  general Newton-Raphson  technique that  solves
the fourth-order (second-order) set  of real equations  and  boundary
conditions  governing   linear,  adiabatic, nonradial (radial)
stellar  pulsations following the   dimensionless formulation 
 of \citet[][]{1971AcA....21..289D}
\citep[see also][]{1989nos..book.....U}.   The
prescription we follow to  assess the run  of the Brunt-V\"ais\"al\"a
frequency ($N$)  is  the so-called ``Ledoux Modified'' treatment
\citep{1990ApJS...72..335T,1991ApJ...367..601B}.

\subsection{Evolutionary sequences}
\label{sequencies}

Realistic  configurations   for low-mass He-core WD and pre-WD stars
were derived by \citet{2013A&A...557A..19A} by  mimicking the binary
evolution  of progenitor stars.  Binary evolution was assumed to be
fully nonconservative, and the loss of angular momentum due to  mass
loss, gravitational wave radiation, and magnetic braking was
considered. All of the He-core pre-WD initial models were derived from
evolutionary calculations for binary systems consisting of an evolving
Main Sequence low-mass component (donor star)  of initially $1
M_{\sun}$ and a $1.4 M_{\sun}$ neutron star companion  as the other
component. A total of 14 initial He-core pre-WD models with stellar
masses between  $0.1554$ and $0.4352 M_{\sun}$ were computed for
initial  orbital periods at the beginning of the Roche lobe phase in
the range $0.9$ to $300$ d.  In this paper,  we focus on the
assessment of the rates of period change values corresponding to the
complete   evolutionary stages of these models  down to the range of
luminosities of cool WDs, including some stages previous to the
thermonuclear CNO flashes during the beginning of the cooling branch.

\begin{figure*}[t]
  \begin{center}
\subfigure{\includegraphics[width=.33\textwidth]{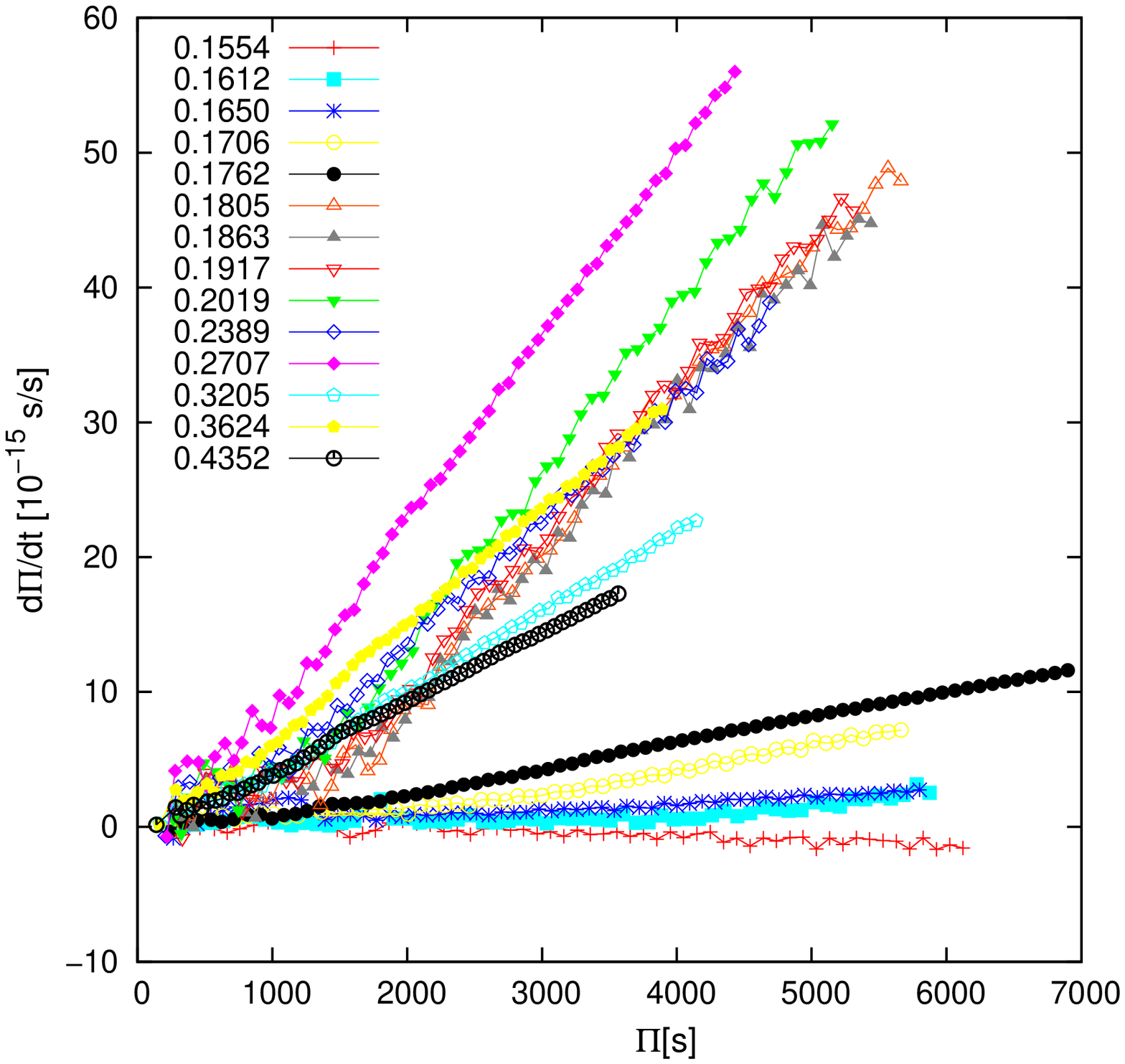}}
\subfigure{\includegraphics[width=.33\textwidth]{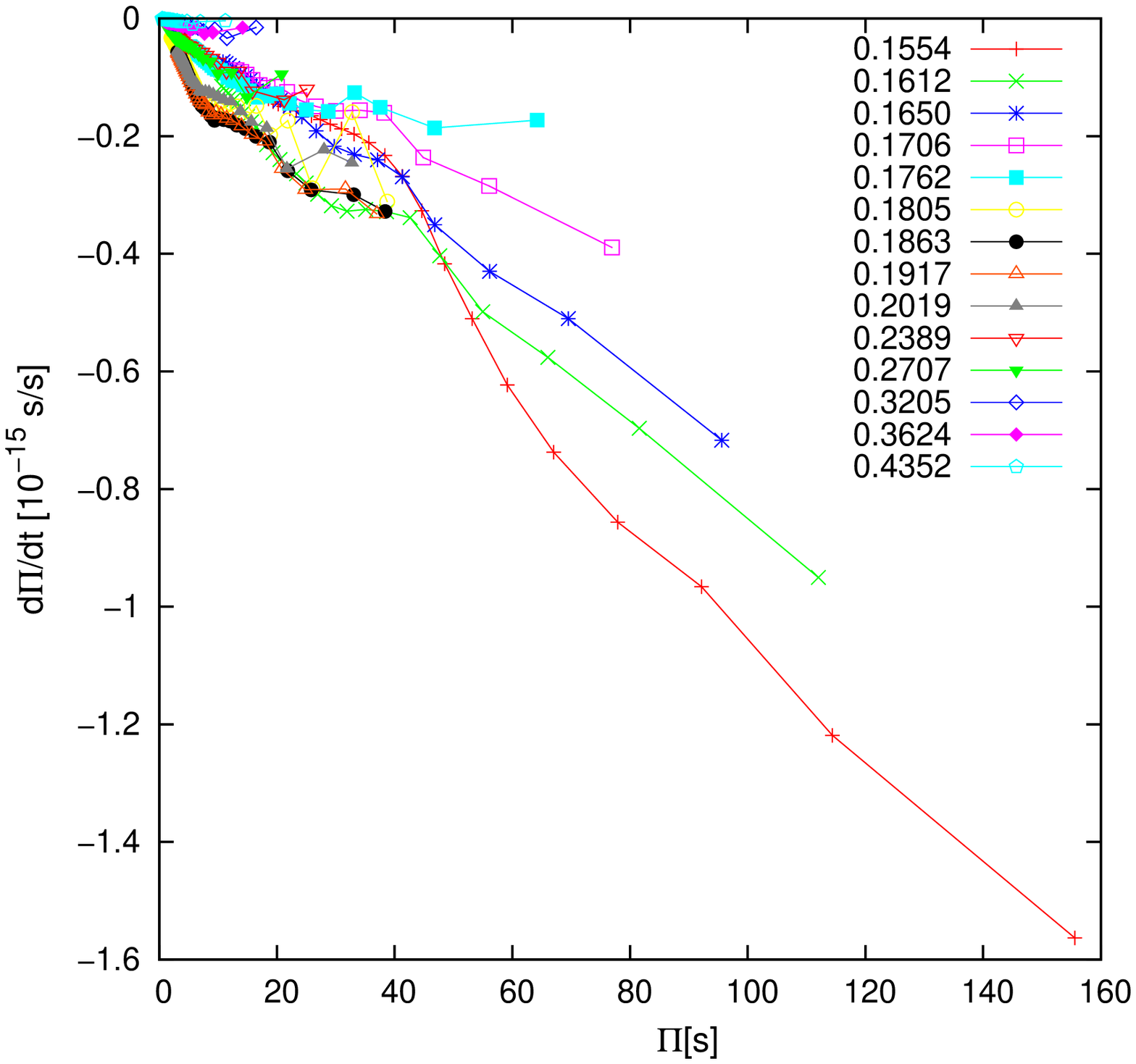}}
\subfigure{\includegraphics[width=.33\textwidth]{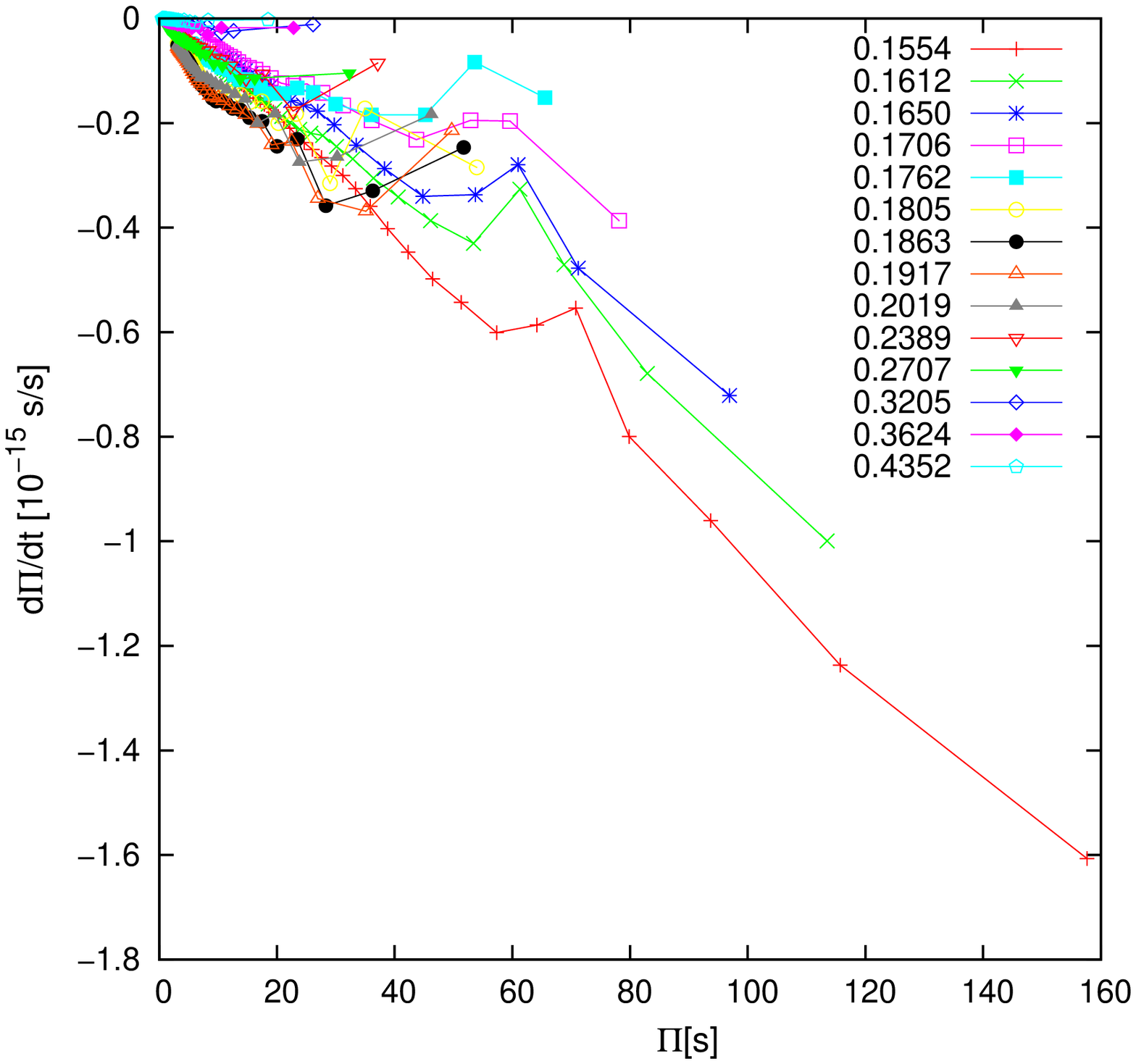}}
  \end{center}
\caption{{\it Left:} The rates of period change of $\ell= 1$ $g$ modes
  versus the pulsation periods, corresponding to WD models
  characterized by an effective temperature of $T_{\rm eff}\sim 9000$ K
  and different stellar masses   
($0.1554 \leq M_{\star}/M_{\sun} \leq 0.4352 $). 
{\it Middle:} Same as in left panel, but for $\ell= 1$ $p$ modes.
{\it Right:} Same as in left panel, but for radial ($\ell= 0$) modes.}
\label{parte1_gl1_pl1_rl0}
\end{figure*}

\section{The theoretical rates of period change}
\label{rates}

In this work, the rates of period change are assessed as simple
differencing of the periods of successive models in each evolutionary
sequence. Specifically, the rate of change of the period $\Pi_k$ at
the time $\tau_i$ is estimated as:

\begin{equation}
\dot{\Pi}_k(\tau_i)= \frac{\Pi_k(\tau_{i})-\Pi_k(\tau_{i-1})}{\Delta \tau_i}, 
\end{equation}

\noindent where $\Delta \tau_i= \tau_{i}-\tau_{i-1}$ is the
evolutionary time step, and $\Pi_k(\tau_{i})$ and $\Pi_k(\tau_{i-1})$
are the pulsation periods of the mode with radial order $k$ evaluated
at the times $\tau_{i}$ and $\tau_{i-1}$, respectively.  In our
computations, the time step $\Delta \tau_i$ is small enough as to
ensure that this simple numerical recipe yields very precise results
for $\dot{\Pi}$. We compute the rate of period change for nonradial
$\ell= 1, 2$ $g$ and $p$ modes, and also radial ($\ell= 0$) modes. The
set  of pulsation modes considered in this work covers a  very wide
range of  periods (up to $\sim 7000$ s), embracing all the
periodicities detected in ELMV and pre-ELMV stars up to now.

Low-mass WDs have real possibilities of being observed at three stages
\citep{2013A&A...557A..19A}: the final cooling branch (WD phase), the
stages at constant luminosity following the end of Roche lobe overflow
(pre-WD phase), and for $M_{\star} \gtrsim 0.18 M_{\sun}$, the
evolutionary stages prior to the occurrence of CNO flashes on the
early cooling branches (pre-flash stages)\footnote{During the rapid
  incursions of the stars in the $\log T_{\rm eff} - \log g$ diagram
  while they are looping between the CNO flashes, the evolution is so
  fast that the probability of finding a star in those stages is
  almost null \citep{2013A&A...557A..19A}. For this reason, we will
  not consider those evolutionary stages in this work.}. Below, we
explore the magnitude and sign of the rates of period changes of
low-mass WD stars at these evolutionary stages. In all the regimes
considered in this work, the rate of period
change values for $\ell= 2$ are of the same order of magnitude
than for $\ell= 1$. Thus, we will concentrate on showing results
only for the case $\ell= 1$, although we must keep in mind that also
modes with $\ell= 2$ can be observed in this type of pulsating stars.

\subsection{WD phase: ELMVs}
\label{wd}

\begin{figure*}[t]
  \begin{center}
\subfigure{\includegraphics[width=.33\textwidth]{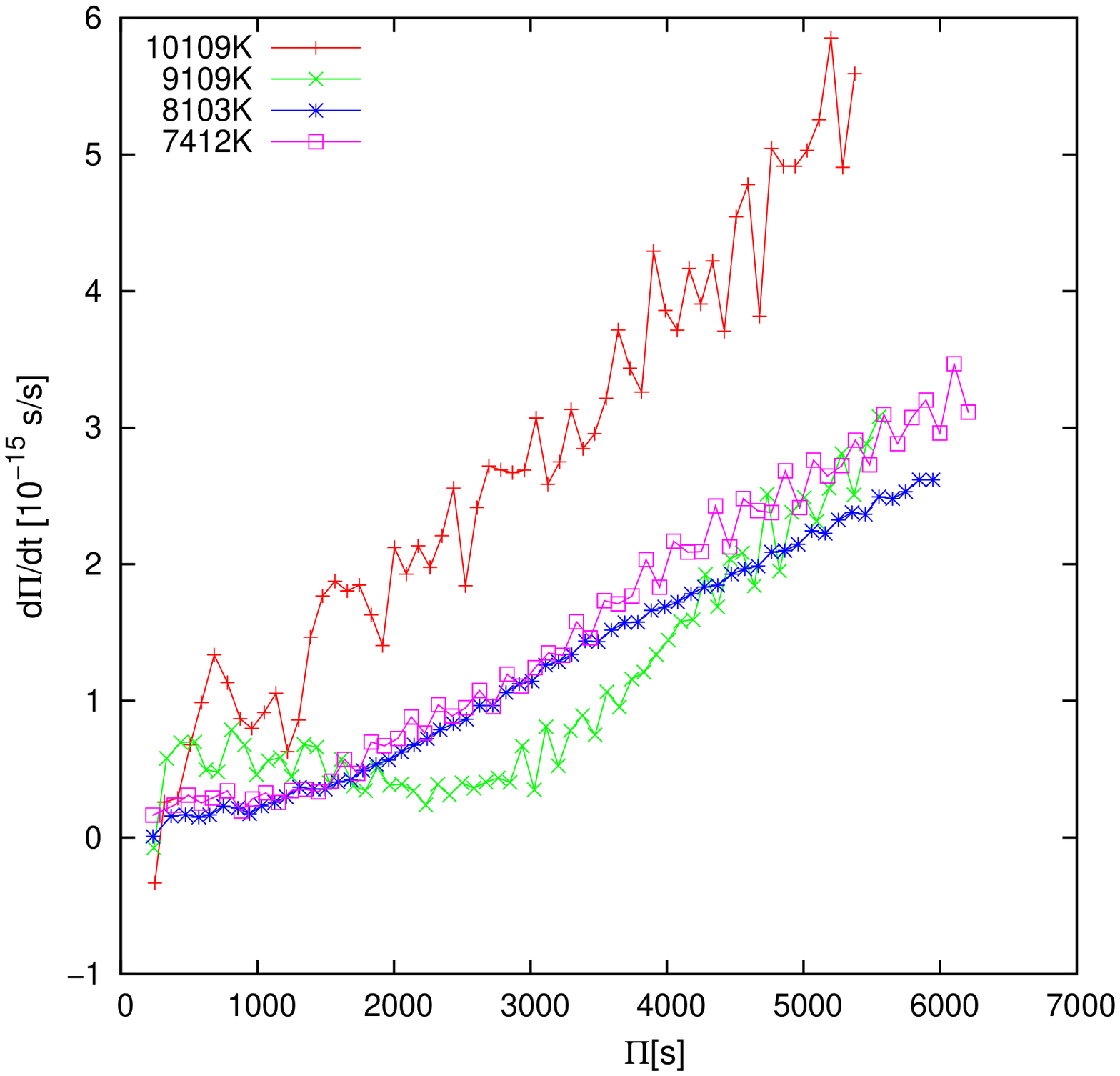}}
\subfigure{\includegraphics[width=.33\textwidth]{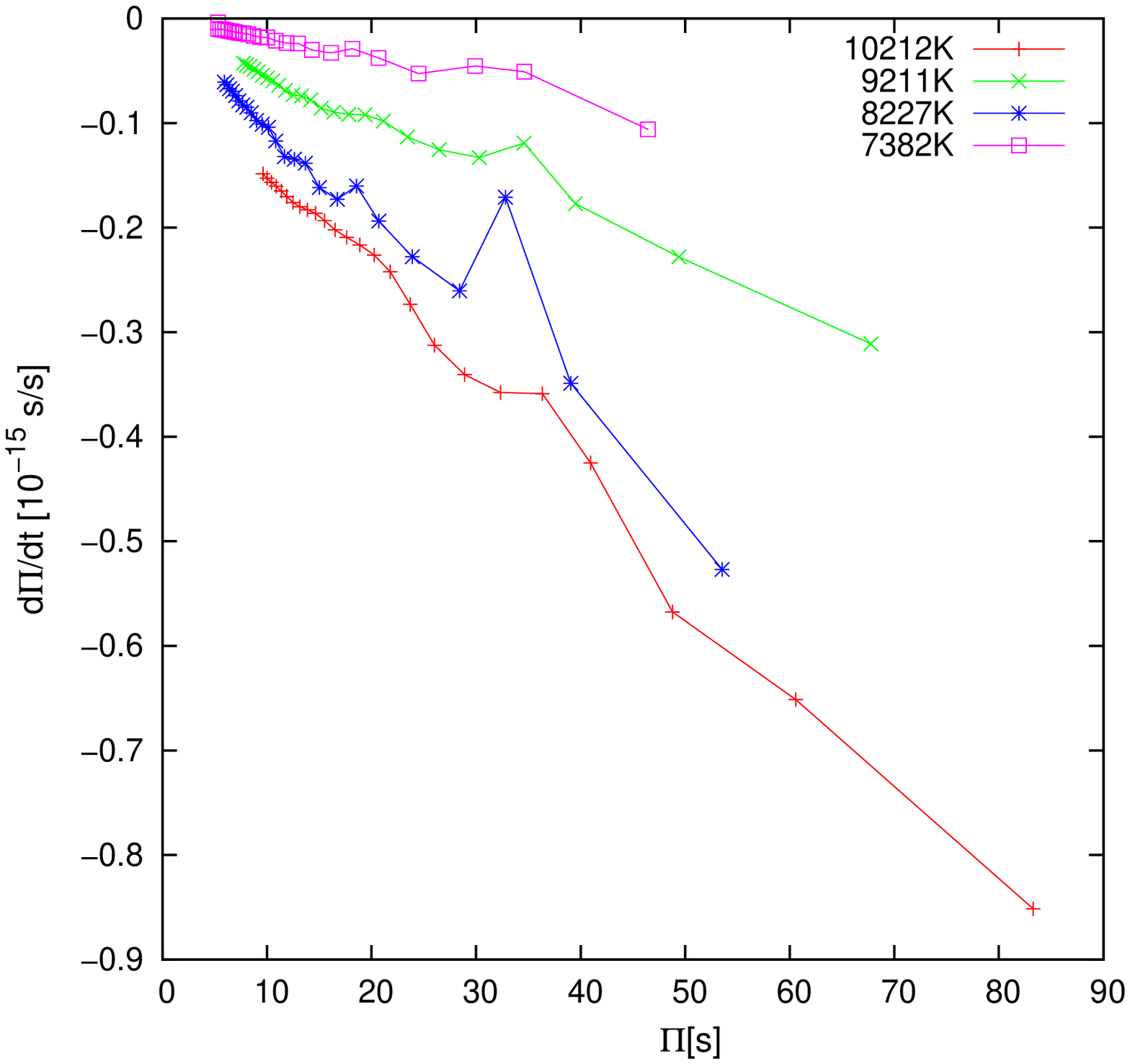}}
\subfigure{\includegraphics[width=.33\textwidth]{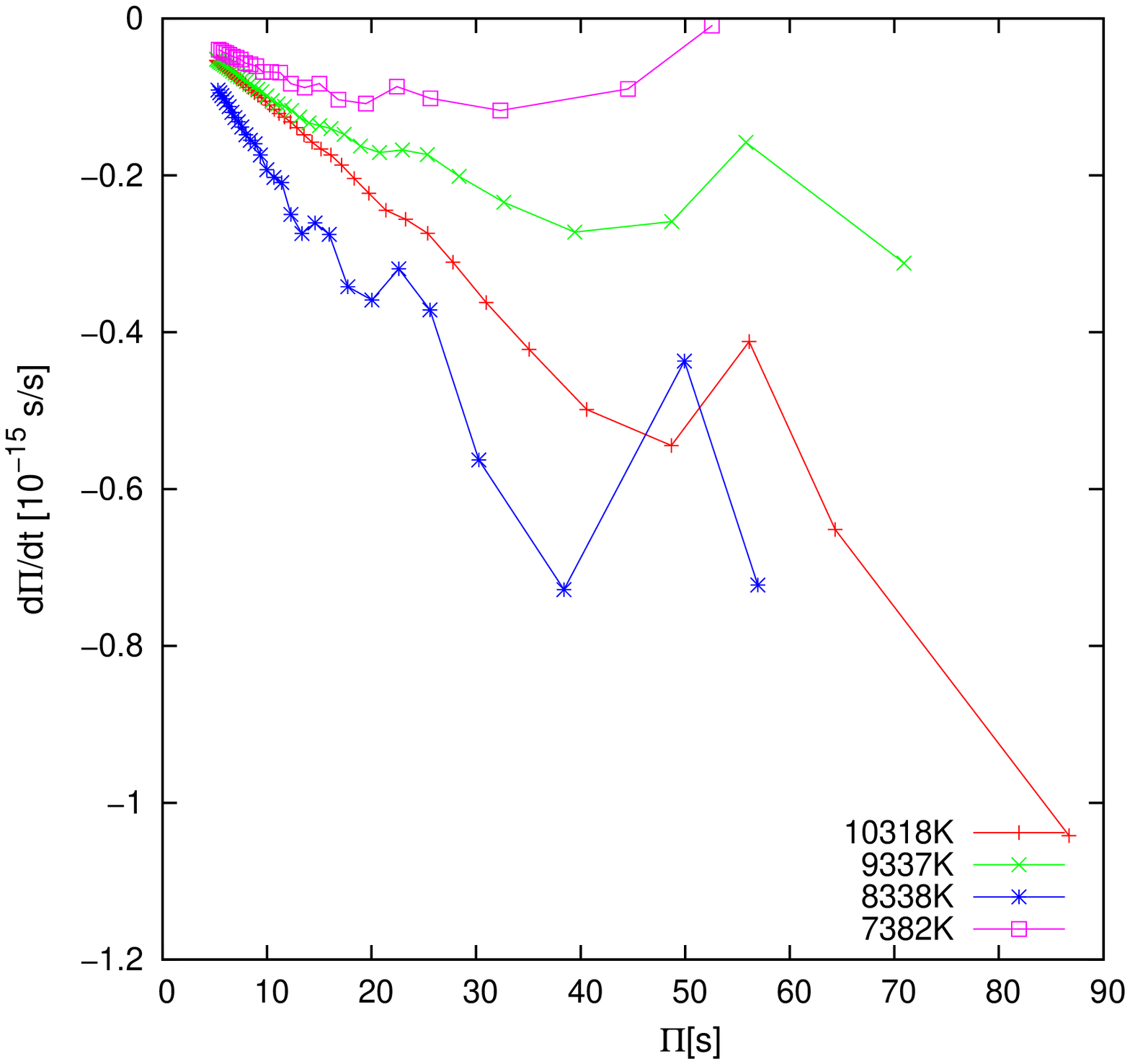}}
  \end{center}
\caption{
{\it Left:} The rate of period change of $\ell= 1$ $g$ modes
  versus the pulsation periods, corresponding to WD models
  characterized by a stellar mass $M_{\star}= 0.1762 M_{\sun}$
  and  different effective temperatures.
{\it Middle:} Similar to left panel,
  but for $\ell= 1$ $p$ modes.
{\it Right:} Similar to left panel,
  but for radial ($\ell= 0$) modes.
}
\label{01762_parte2_gl1_pl1_rl0}
\end{figure*}

\begin{figure*}[t]
  \begin{center}
\subfigure{\includegraphics[width=.33\textwidth]{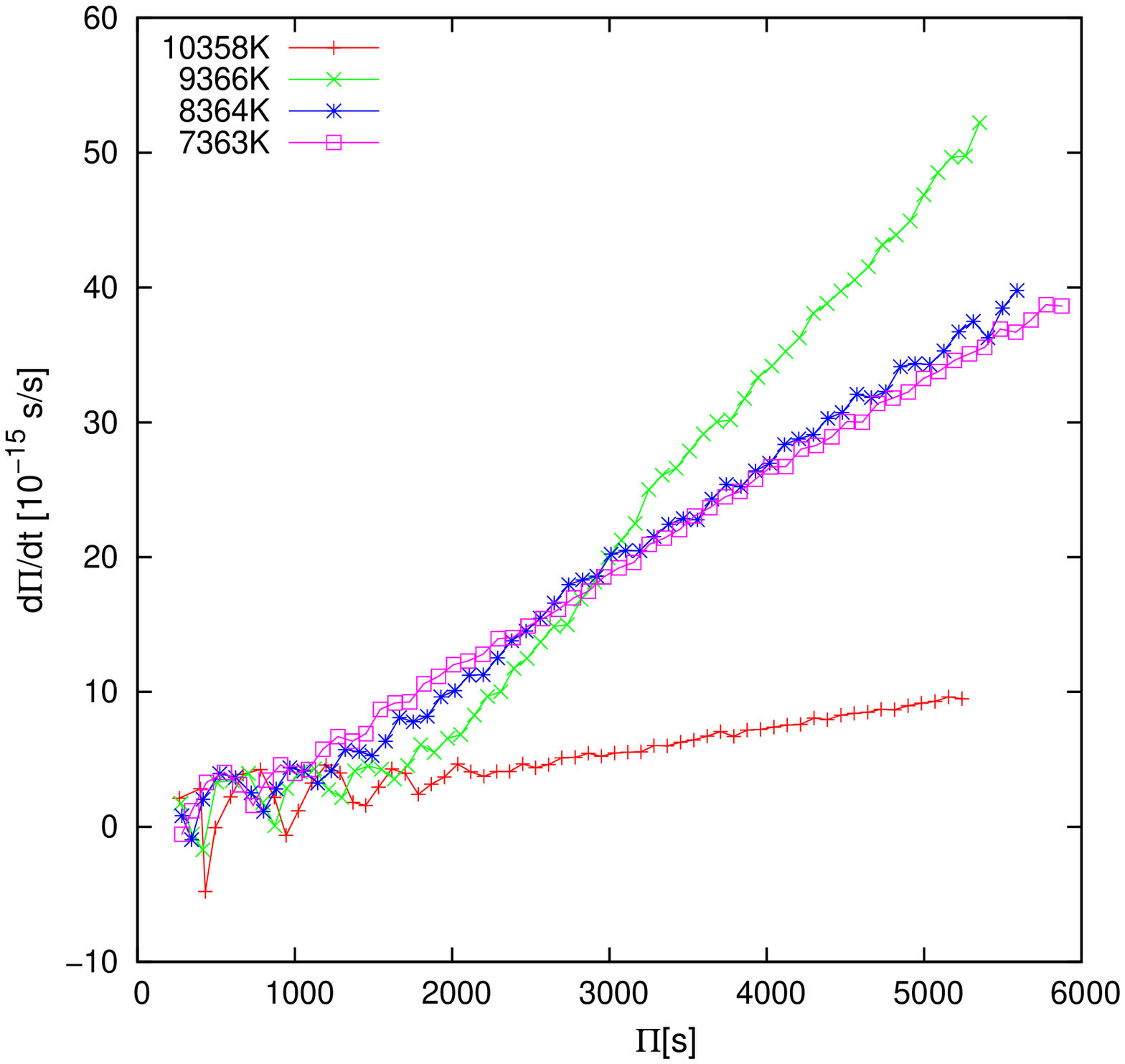}}
\subfigure{\includegraphics[width=.33\textwidth]{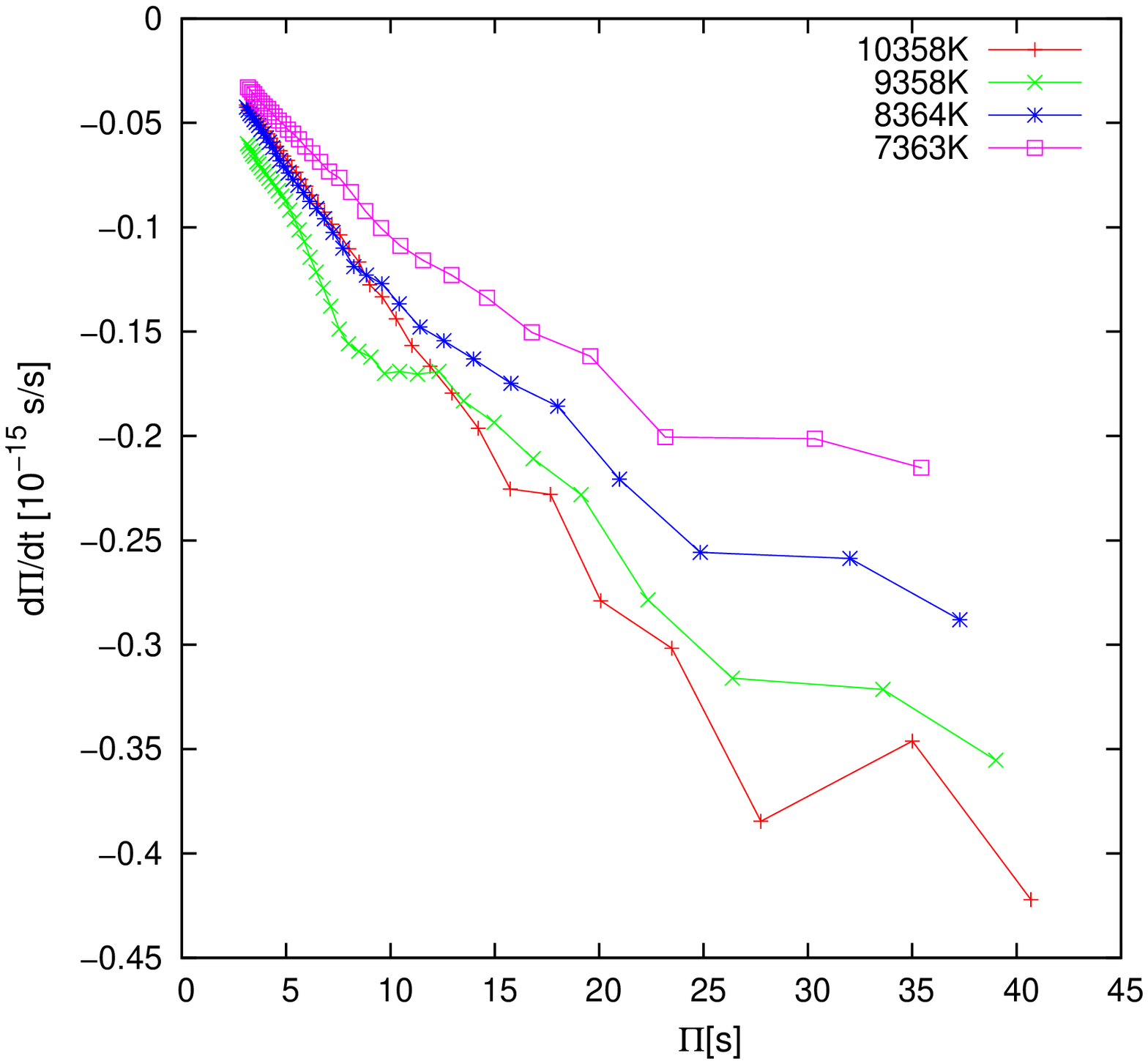}}
\subfigure{\includegraphics[width=.33\textwidth]{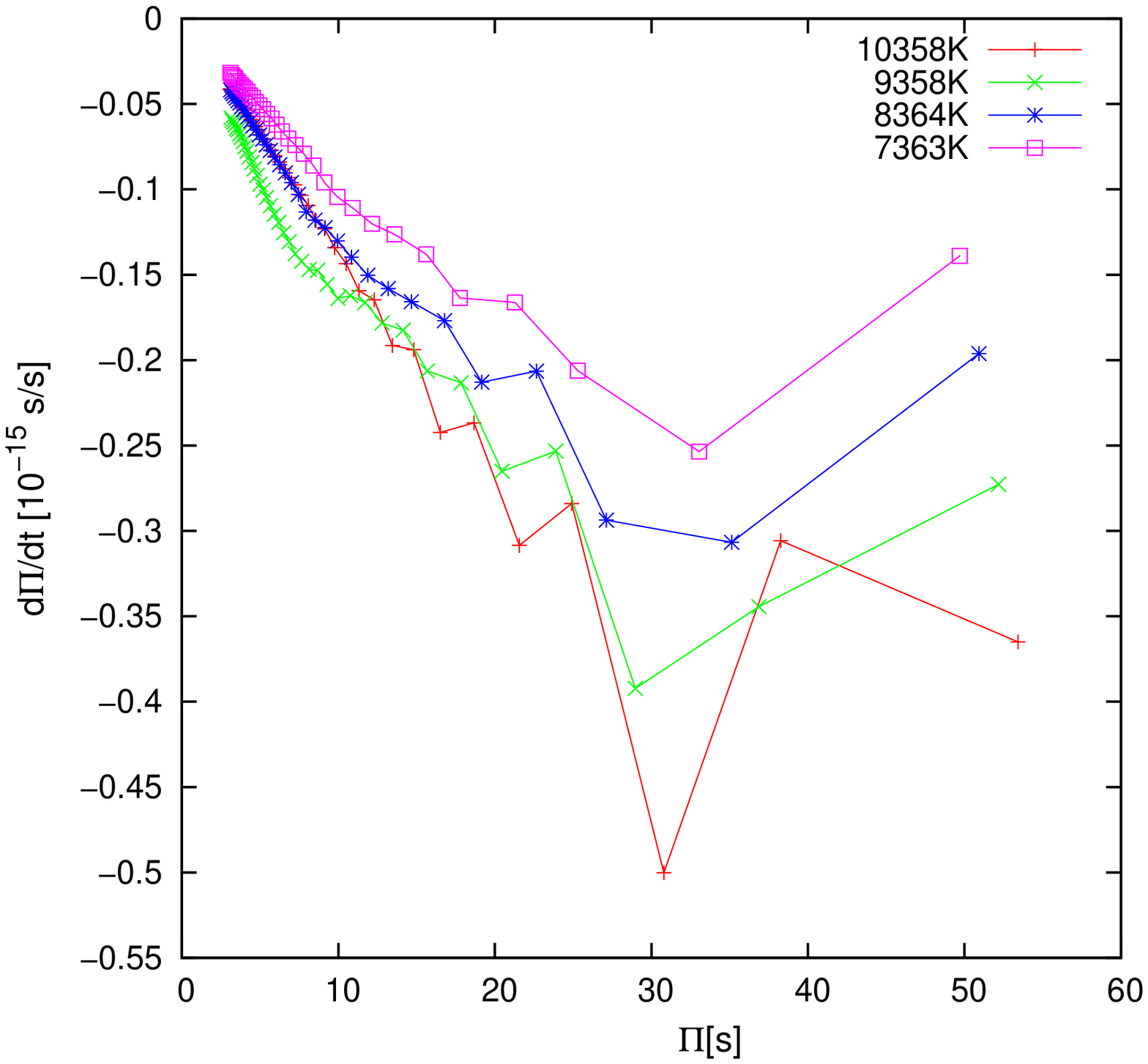}}
  \end{center}
\caption{
{\it Left:} The rate of period change of $\ell= 1$ $g$ modes
  versus the pulsation periods, corresponding to WD models
  characterized by a stellar mass $M_{\star}= 0.1863 M_{\sun}$
  and  different effective temperatures.
{\it Middle:} Similar to the left panel, but for $\ell= 1$ $p$ modes.
{\it Right:} Similar to the left panel, but for radial ($\ell= 0$) modes.
}
\label{01863_parte2_gl1_pl1_rl0}
\end{figure*}

Next, we shall examine the effect of changing the stellar mass and the
effective temperature on the rate of period change of ELMV models,
that is, low-mass WD models already evolving in their final cooling
branches. The adiabatic and nonadiabatic pulsation   properties of
pulsating low-mass WD stars, or ELMVs, have been explored in detail in
\citet{2012A&A...547A..96C,2013ApJ...762...57V,2014A&A...569A.106C,2016A&A...585A...1C}. In
Fig. \ref{HR-ELMs} we present a $T_{\rm eff} - \log g$ diagram showing
the low-mass He-core WD evolutionary tracks of
\citet{2013A&A...557A..19A}, with the stellar masses indicated with
small numbers. For illustrative purposes, we also include the
location of the ten known ELMVs \citep{2012ApJ...750L..28H,
  2013ApJ...765..102H,2013MNRAS.436.3573H,2015MNRAS.446L..26K,2015ASPC..493..217B,2016arXiv161206390B}
with red circles, and stars observed  not to vary
\citep{2012PASP..124....1S,2012ApJ...750L..28H,2013ApJ...765..102H,2013MNRAS.436.3573H},
displayed with small black circles. The gray squares (triangles) on
the evolutionary tracks indicate the location of the template models
to be analyzed in Fig. \ref{parte1_gl1_pl1_rl0}
(Figs. \ref{01762_parte2_gl1_pl1_rl0} and
\ref{01863_parte2_gl1_pl1_rl0}) below. 

\begin{figure*}[t]
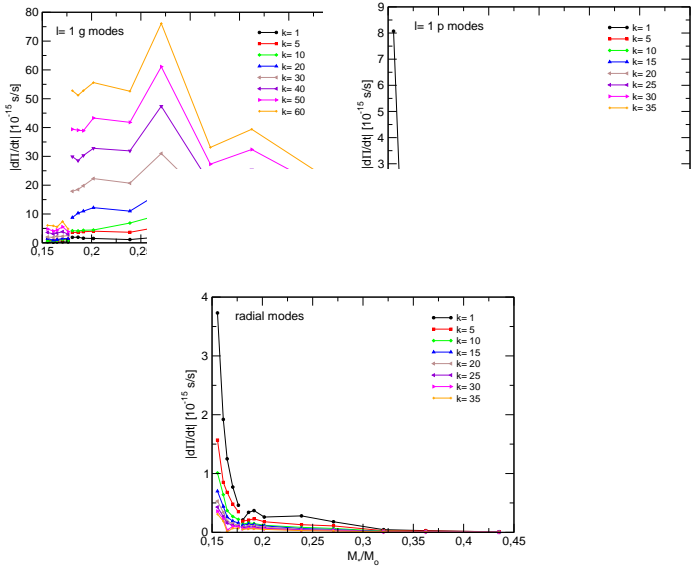

  \begin{center}
\subfigure{\includegraphics[width=.33\textwidth]{max-g.eps}}
\subfigure{\includegraphics[width=.33\textwidth]{max-p.eps}}
\subfigure{\includegraphics[width=.33\textwidth]{max-r.eps}}
  \end{center}
\caption{
{\it Left:} The absolute value of the maximum rate of period
  change in terms of the stellar mass corresponding
  to $\ell= 1$ $g$ modes with selected radial orders $k$,
  for WD models with effective temperatures in the range
  $8000 \lesssim T_{\rm eff} \lesssim 10000$ K.
{\it Middle:} Same as in the left panel, 
  but for $\ell= 1$ $p$ modes.
{\it Right:} Same as in the left panel, 
 but for radial ($\ell= 0$) modes.
}
\label{max_rate_g_p_r}
\end{figure*}

In assessing the dependence of $\dot{\Pi}$ for $g$ modes with
$M_{\star}$ and $T_{\rm eff}$ in WD stars, it is usual to consider the
predictions of the simple cooling model of \citet{1952MNRAS.112..583M}
for comparison with numerical results.  Within the framework  of the
Mestel's  cooling law, \citet {1986ApJ...302..530K} have  derived a
relation  between $\dot{\Pi}$ and $M_{\star}$ and $T_{\rm eff}$ (their
Eq. 3), which predicts that $\dot{\Pi}$ is larger  for lower stellar
masses and higher effective temperatures. The mass  dependence can be
understood by realizing that the lower the mass, the larger the
radiating surface and the lower the total heat capacity.  For a fixed
$T_{\rm eff}$ value, less massive models have higher luminosities  and thus cool
faster with a larger $\dot{\Pi}$. Concerning the dependence  of
$\dot{\Pi}$ with the effective temperature, for a fixed $M_{\star}$,
models with higher $T_{\rm eff}$ cool faster, with the consequence
that $\dot{\Pi}$ is larger. This simple picture becomes more
complicated when there exist another energy source like nuclear
burning, apart from the heat reservoir stored in the ions during the
previous evolutionary phases.  This is the case of low-mass WDs with
$M_{\star} \lesssim 0.18-0.20 M_{\sun}$ (ELM WDs), which are
characterized by intense H burning. Thus, the simple predictions based
on the Mestel theory cannot be applied to WDs in this mass range.

We depict in Fig. \ref{parte1_gl1_pl1_rl0} the rates of period change
in terms of periods for $\ell= 1$ $g$ modes (left panel), $\ell= 1$
$p$ modes (middle panel), and $\ell= 0$ radial modes (right panel),
for low-mass WD models with $T_{\rm eff} \sim 9000$ K and several
stellar masses ($0.1554 \leq M_{\star}/M_{\sun} \leq 0.4352 $). The
location of these stellar models in the $T_{\rm eff} - \log g$ diagram
is marked with gray squares in Fig. \ref{HR-ELMs}.  For $g$ modes
(left panel of Fig. \ref{parte1_gl1_pl1_rl0}), $\dot{\Pi}$ linearly
increases with the radial order $k$ and thus with period, which is the
reflection of the increase of $\Pi$ with $k$. Two well distinguishable
branches of the $\dot{\Pi}$ vs $\Pi$ relationship are visible in the
figure, one of them corresponding to models with $M_{\star} \gtrsim
0.18 M_{\sun}$, and the other one associated to models with $M_{\star}
\lesssim 0.18 M_{\sun}$. In the first group of models,  nuclear
burning is not relevant, and so, the rate of period change is
generally larger for lower stellar mass (at fixed $T_{\rm eff} \sim
9000$ K) and all the $\dot{\Pi}$ values are positive. Note that for
this set of models, $g$ modes are very sensitive to the He/H
composition gradient \citep[see Fig. 8 of][]{2014A&A...569A.106C}.
The fact that  $\dot{\Pi} > 0$ implies that the rates of period change
in models with $M_{\star} \gtrsim 0.18 M_{\sun}$ are dominated by
cooling (first term in Eq. \ref{eq-dotp}). The rates of period change
for this set of models range from $\sim 10^{-15}$ s/s for the shortest
$g$-mode periods\footnote{Note that the $\dot{\Pi}$ value for some
  low-order modes is very close to zero, or even negative.}  up to
$\sim 7 \times 10^{-14}$ s/s for $\Pi \sim 6000$ s.

\begin{table*}
\centering
\caption{The absolute value of the upper limit of the theoretical
  rates of period change,
  $|\dot{\Pi}_{\rm max}|$ (in units of $10^{-15}$ s/s), for
  selected $\ell= 1$ $g$ modes, corresponding to
  low-mass WD models with effective temperatures in the range
  $8000 \lesssim T_{\rm eff} \lesssim 10\,000$ K.
  A graphical representation of these values
  is shown in the left panel of Fig. \ref{max_rate_g_p_r}.}
\begin{tabular}{ccccccccc}
\hline
\hline
\noalign{\smallskip}
$M_{\star}/M_{\sun}$  & $k= 1$ & $k= 5$ & $k= 10$ & $k= 20$ & 
$k= 30$ & $k= 40$ & $k= 50$ & $k= 60$ \\
\noalign{\smallskip}
\hline
\noalign{\smallskip}
0.1554 & 0.35 & 0.75 & 0.63 & 1.50 & 2.00 & 3.60 & 4.80 & 6.05 \\
0.1612 & 0.20 & 0.85 & 0.69 & 0.91 & 1.90 & 3.10 & 4.10 & 5.90 \\
0.1650 & 0.28 & 0.93 & 0.81 & 1.08 & 2.27 & 3.39 & 4.45 & 5.52 \\
0.1706 & 0.32 & 1.07 & 1.18 & 1.54 & 2.25 & 3.85 & 5.56 & 7.42 \\
0.1762 & 0.30 & 1.17 & 0.88 & 1.43 & 2.58 & 3.00 & 3.90 & 4.82 \\
\noalign{\smallskip}
\hline
\noalign{\smallskip}
0.1805 & 1.90 & 3.61 & 4.20 & 8.77 & 17.9 & 29.9 & 39.4 & 52.8 \\
0.1863 & 1.95 & 3.63 & 4.14 & 10.3 & 18.5 & 28.5 & 39.1 & 51.2 \\
0.1917 & 1.60 & 3.84 & 4.37 & 11.0 & 19.8 & 30.3 & 38.9 & 52.8 \\
0.2019 & 1.51 & 4.05 & 4.45 & 12.2 & 22.3 & 32.8 & 43.3 & 55.6 \\
0.2389 & 1.13 & 3.67 & 6.87 & 11.0 & 20.7 & 31.9 & 41.8 & 52.6 \\
0.2707 & 2.10 & 5.65 & 9.86 & 17.7 & 31.0 & 47.4 & 61.1 & 76.1 \\
0.3205 & 0.15 & 2.55 & 3.70 & 8.06 & 14.2 & 20.4 & 27.3 & 33.1 \\
0.3624 & 0.16 & 3.11 & 4.78 & 10.1 & 17.7 & 25.3 & 32.4 & 39.4 \\
0.4352 & 0.19 & 2.00 & 3.05 & 6.24 & 10.4 & 14.7 & 19.0 & 23.2 \\
\hline
\hline
\end{tabular}
\label{table1}
\end{table*}

The $\dot{\Pi}$ values for the group of models with  $M_{\star}
\lesssim 0.18 M_{\sun}$, on the other hand, are lower than $\sim
10^{-14}$ s/s, and are indeed substantially smaller than for the first
group of models. This is due to the fact that, for models with stellar
masses lower than the threshold mass of $\sim 0.18 M_{\sun}$, the
evolution is dominated by nuclear burning. As a result, the WD cooling
is markedly delayed, in such a way that the rates of period change are
smaller in magnitude as compared with the case in which nuclear
burning is negligible ($M_{\star} \gtrsim 0.18 M_{\sun}$).  Note that
$g$ modes in this mass range mainly probe the core regions \citep[see
  Fig. 7 of][]{2014A&A...569A.106C}.  The $\dot{\Pi}$ values for
models $M_{\star} \lesssim 0.18 M_{\sun}$ are smaller for lower
stellar mass, as it can be seen in left panel of
Fig. \ref{parte1_gl1_pl1_rl0}.  This trend is opposite to that
predicted by the simple formula of \citet {1986ApJ...302..530K} (see
above). The fact that in this mass range
($0.15 \lesssim M_{\star}/M_{\sun} \lesssim 0.18$) the lowest-mass models
are characterized by smaller $\dot{\Pi}$ values is due to that
these models have more intense nuclear burning, which implies smaller cooling
timescales. In this context, it is expected that  the first term in
Eq. (\ref{eq-dotp}) (cooling) for these sequences must be small, and
the second term in this equation (gravitational contraction) must be
large, as compared with the case of the most massive WD sequences
($M_{\star} \gtrsim 0.18 M_{\sun}$). This is precisely what  we found
by examining the evolution of the stellar radius in our evolutionary
sequences.  Interestingly enough, for the lowest-mass  model
($M_{\star}= 0.1554 M_{\sun}$), most of the $\dot{\Pi}$ values are
negative.  This means that the pulsation periods of  $g$ modes for
this model generally shorten while the WD cools down. This novel
result can be understood by realising that, for this WD sequence, the
gravitational contraction is so large that the term in $\dot{T}/T$ of
Eq. (\ref{eq-dotp}) overcomes the term in $\dot{R_{\star}}/R_{\star}$
and so, the rates of period change are negative.

We now examine the trend of the rates of period change with the
pulsation periods for $p$ modes and radial modes, corresponding to the
same template models at $T_{\rm eff} \sim 9000$ K.  According to the
results of \citet[][]{2014A&A...569A.106C}, $p$- and radial-mode
periods decrease with decreasing $T_{\rm eff}$ (see their Fig. 19).
Therefore, we expect the $\dot{\Pi}$ values for these kinds of modes
 to  be all negative. This expectation is borne out by examining the middle
and right panels of Fig. \ref{parte1_gl1_pl1_rl0}. In the case of $p$
modes, this behaviour is due to the fact that
the Lamb frequency ---which is the
critical frequency of nonradial $p$ modes--- increases for decreasing
effective temperatures \citep{2014A&A...569A.106C}, in such a way that
the whole $p$-mode frequency spectrum shifts to higher frequencies,
and so to shorter periods, as the WD cools. Radial modes exhibit the
same behaviour as $p$ modes.  Regarding the dependence of $\dot{\Pi}$
with the stellar mass, according to our numerical results, the
magnitude of the rates of period change is larger for lower
$M_{\star}$, as shown in Fig. \ref{parte1_gl1_pl1_rl0}.  Note that, in
this case, there is no obvious  difference in the behaviour of
$\dot{\Pi}$ between models with intense nuclear burning ($M_{\star}
\lesssim 0.18 M_{\sun}$) or models characterized by negligible nuclear
burning ($M_{\star} \gtrsim 0.18 M_{\sun}$). All in all, the
magnitudes of the rates of period change are smaller than $\sim 1.6
\times 10^{-15}$ s/s for the range of radial- and $p$-mode periods
analyzed. These $\dot{\Pi}$ values are an order of magnitude smaller
than for $g$ modes, and then, they would be very difficult (if not
impossible) to measure in the eventual case in which radial modes
and/or nonradial $p$ modes were detected in low-mass WDs. It is worth
noting that \citet{2013ApJ...765..102H} have found short periods
($\sim 100-140$ s) in the light curve of the ELMV star SDSS
J111215.82$+$111745.0 that could be associated to $p$ modes (or even
radial modes), but these observations need confirmation.

\begin{figure} 
\begin{center}
\includegraphics[clip,width=9 cm]{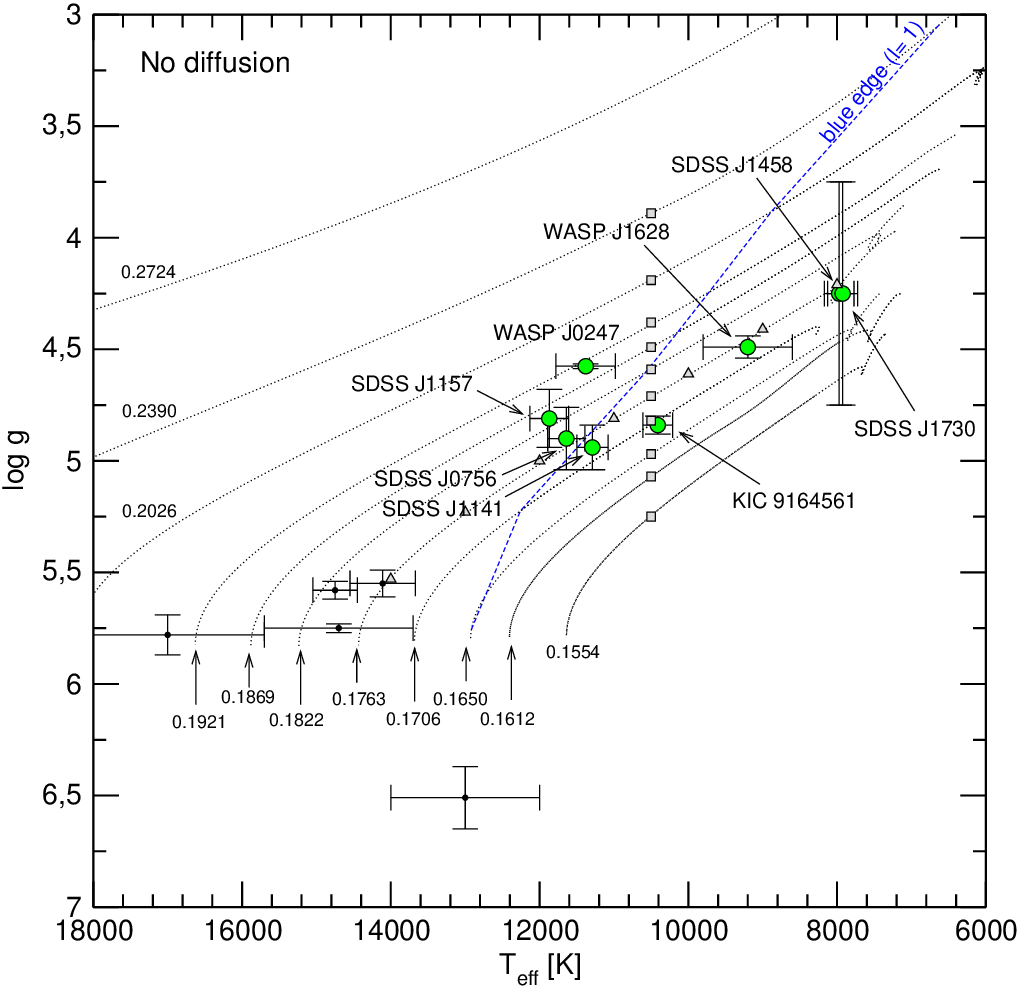} 
\caption{$T_{\rm eff} - \log g$ diagram showing our low-mass He-core 
  pre-WD evolutionary tracks (thin dotted black lines) computed
  neglecting element 
 diffusion. Numbers correspond to the stellar mass of each sequence. 
   Green circles with error bars correspond to the known 
   pre-ELMV stars \citep{2013Natur.498..463M,2014MNRAS.444..208M,2016A&A...587L...5C,2016ApJ...822L..27G,2016ApJ...821L..32Z}, and small black circles
  depict the location of pre-ELM (nonvariable) WDs 
  observed in the \emph{Kepler} mission field \citep{2015ApJ...803...82R,2010ApJ...715...51V,2011ApJ...728..139C,2012ApJ...748..115B}.
  The dashed blue line indicates the nonradial
  dipole ($\ell= 1$)  blue edge of the pre-ELMV  
  instability domain  due to the $\kappa-\gamma$ mechanism acting
  at the He$^+-$He$^{++}$ partial 
  ionization   region, as obtained in \citet{2016A&A...588A..74C}.
  The gray squares and triangles on the evolutionary tracks indicate the 
 location of the template models analyzed in the text.}
\label{HR-pre-ELMs} 
\end{center}
\end{figure}

In the left panels of Figs. \ref{01762_parte2_gl1_pl1_rl0} and
\ref{01863_parte2_gl1_pl1_rl0}  we display the rate of period change
of $\ell= 1$ $g$ modes versus the pulsation periods, corresponding to
template WD models characterized by stellar masses $M_{\star}= 0.1762
M_{\sun}$ and $M_{\star}= 0.1863 M_{\sun}$, respectively,  and
different effective temperatures. Middle and right panels of the same
figure correspond to $p$- and radial-mode periods, respectively.  The
location of these template models in the $T_{\rm eff} - \log g$ plane
is indicated with gray triangle symbols on the pertinent evolutionary
tracks of Fig. \ref{HR-ELMs}. We first focus on the results for $g$
modes of the $0.1762 M_{\sun}$ sequence at $T_{\rm eff} \sim 10100$ K,
9100 K, 8100 K and $7400$ K (left panel of
Fig. \ref{01762_parte2_gl1_pl1_rl0}). Unlike what might be expected,
the $\dot{\Pi}$ values do not vary monotonously with the effective
temperature, at least for long periods ($\gtrsim 1500$ s). Indeed,
$\dot{\Pi}$ first decreases from $T_{\rm eff} \sim 10100$ K to $T_{\rm
  eff} \sim 9100$ K, and then increases for lower $T_{\rm eff}$.  The
rates of period change are all positive for the complete range of
effective temperatures considered, except in the case of the $k= 1$
$g$ mode, which exhibit $\dot{\Pi} < 0$ for the hottest template
models. For this sequence of models and for the complete range of
$T_{\rm eff}$, we found $\dot{\Pi} \lesssim 6 \times 10^{-15}$
s/s. For $p$ and radial modes, the rates of period change are all
negative, and their absolute values are below $\dot{\Pi} \lesssim 9
\times 10^{-16}$ s/s (middle and right panel of
Fig. \ref{01762_parte2_gl1_pl1_rl0}). Similar to what happens for $g$
modes, for radial and $p$ modes the magnitude of the rates of period
change do not monotonously change with $T_{\rm eff}$, although on
average, $|\dot{\Pi}|$ decreases with decreasing effective
temperature.

\begin{figure} 
\begin{center}
\includegraphics[clip,width=9 cm]{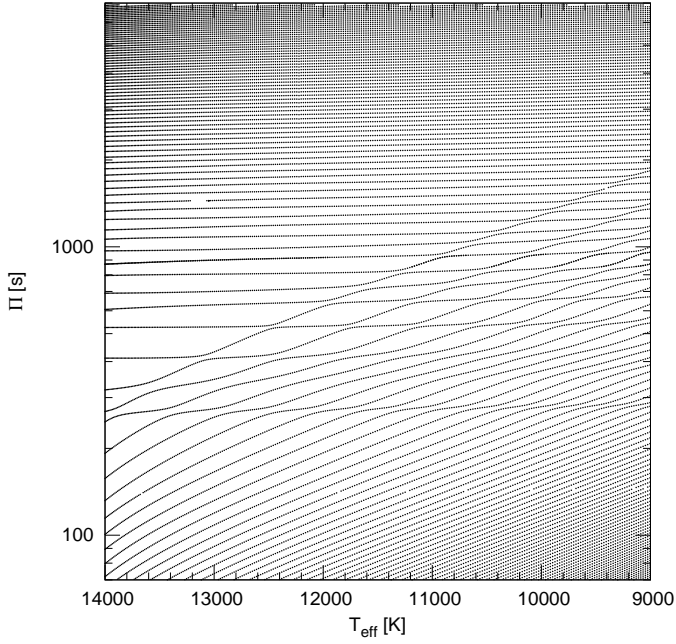} 
\caption{Evolution of the pulsation periods of $p$, $g$, and $p-g$
  mixed modes with $\ell= 1$ from low to high $T_{\rm eff}$ values
  for pre-WD models
  with $M_{\star}= 0.1763 M_{\sun}$. Note the multiple avoided crossings
experienced for the modes with periods in the range $250-2000$ s.}
\label{per-teff-01763} 
\end{center}
\end{figure}

The behaviour of $\dot{\Pi}$ vs $\Pi$
found for models with $M_{\star}= 0.1863 M_{\sun}$
(Fig. \ref{01863_parte2_gl1_pl1_rl0}) is qualitatively similar to
that described for the models with  $M_{\star}= 0.1762 M_{\sun}$.  
The only relevant difference is that the magnitude of the rate
of period change for $g$ modes (left panel) is notoriously greater
than for $M_{\star}= 0.1762 M_{\sun}$. This is expected because,
as stated before in this paper and in the previous papers of this series,
models with $M_{\star} \gtrsim 0.18 M_{\sun}$ do not experience
appreciable nuclear burning and evolve much faster than models
with $M_{\star} \lesssim 0.18 M_{\sun}$. On the other hand, 
the $|\dot{\Pi}|$ values of $p$ and radial modes for the models
with $M_{\star}=  0.1863 M_{\sun}$ are smaller (about a half) than
those corresponding to models with  $M_{\star}= 0.1762 M_{\sun}$.
According to our results for $M_{\star}=  0.1863 M_{\sun}$, 
$\dot{\Pi} \lesssim 6 \times 10^{-14}$ s/s for $g$ modes, and 
$|\dot{\Pi}| \lesssim 5 \times 10^{-16}$ s/s for $p$ and radial
modes (Fig. \ref{01863_parte2_gl1_pl1_rl0}) in the
range of periods considered ($\Pi \lesssim 6000$ s).

\begin{figure} 
\begin{center}
\includegraphics[clip,width=9 cm]{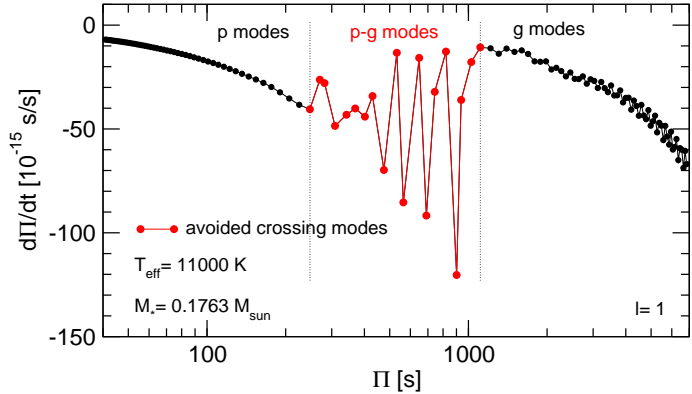} 
\caption{$\dot{\Pi}$ vs $\Pi$ of $\ell= 1$ modes for a
  $0.1763 M_{\sun}$ pre-WD model at $T_{\rm eff}= 11\,000$ K.
  $p-g$ mixed modes that take part in avoided crossings are emphasized
  with red dots.}
\label{per-teff-01763-11000} 
\end{center}
\end{figure}

We close this section by considering the absolute value of the maximum
rate of period change that is theoretically expected in pulsating low-mass
He-core WDs (ELMVs) for the complete interval of effective temperature
and mass range considered in this work.  We display in
Fig. \ref{max_rate_g_p_r} the $|\dot{\Pi}|_{\rm max}$ values in terms
of $M_{\star}$ for $\ell= 1$ $g$ modes (left panel), $p$ modes (middle
panel) and radial modes (right panel) for selected values
of the radial order $k$, covering the range of periods observed
in ELMV stars. In the case of $g$ modes, there
is a clear distinction in the magnitude of $|\dot{\Pi}|_{\rm max}$
depending  on whether $M_{\star} \gtrsim 0.18 M_{\sun}$ or $M_{\star}
\lesssim 0.18 M_{\sun}$. Indeed, we expect to measure much larger
values of $\dot{\Pi}$ (up to $8 \times 10^{-14}$ s/s) for WD stars
with masses larger than $\sim 0.18 M_{\sun}$ than for stars with
$M_{\star} \lesssim 0.18 M_{\sun}$ ($\dot{\Pi} \sim 8 \times 10^{-15}$
s/s at most). We show in Table \ref{table1} the maximum
absolute values of the rates of period change for $g$ modes
for selected values of the radial order $k$ corresponding
to each stellar mass considered in this work (left panel of
Fig. \ref{max_rate_g_p_r}). In the case of $p$ and radial modes
(middle and right 
panels of Fig. \ref{max_rate_g_p_r}), our models predict magnitudes of
the rates of period changes much smaller, up to $\sim 8 \times
10^{-15}$ s/s, corresponding to the lowest-order modes and the
lowest-mass WDs ($M_{\star} \sim 0.15 M_{\sun}$). Note that this range
of $\dot{\Pi}$ is similar to that of the $g$ modes for
$M_{\star} \lesssim 0.18 M_{\sun}$, although for $p$
and radial modes it is expected that $\dot{\Pi} < 0$ for all the
modes. 

\subsection{Pre-WD phase: pre-ELMVs}
\label{pre-wd}

Here, we concentrate on the theoretical rates of period change for
stellar models at evolutionary stages previous to the WD stage, that
is, before the stars reach their maximum effective temperature at the
beginning of the first cooling branch.  As described in
\citet{2013A&A...557A..19A}, models with masses higher than $\sim 0.18
M_{\sun}$ experience multiple CNO flashes after this pre-WD stage and
before enter their final cooling tracks. The nonadiabatic properties
of pulsating low-mass pre-WD stars, or pre-ELMVs, have been explored
at length in
\citet{2016A&A...588A..74C,2016ApJ...822L..27G,2016A&A...595L..12I}.
In Fig. \ref{HR-pre-ELMs} we present a $T_{\rm eff} - \log g$ diagram
showing our low-mass He-core pre-WD evolutionary tracks (thin dotted
black curves) without element diffusion\footnote{We have chosen to
  explore the case in which element diffusion is not allowed to
  operate, due to that recent nonadiabatic studies indicate that
  element diffusion must be weakened so that pulsations can be excited
  by the $\kappa$ mechanism acting at the zone of the second partial
  ionization of He \citep{2016A&A...588A..74C,2016A&A...595L..12I}.
  Exploratory computations in which element diffusion is considered do
  not show substantial differences regarding the $\dot{\Pi}$ values
  reported in this work.}. The stellar mass of each sequence is
indicated with a small number.  For illustrative purposes, we show the
location of the known pre-ELMVs
\citep{2013Natur.498..463M,2014MNRAS.444..208M,2016A&A...587L...5C,
  2016ApJ...822L..27G,2016ApJ...821L..32Z}, and the stars observed not
to vary (small black circles). The gray squares and triangles indicate
the location of the template models to be analyzed below. 

\begin{figure*}[t]
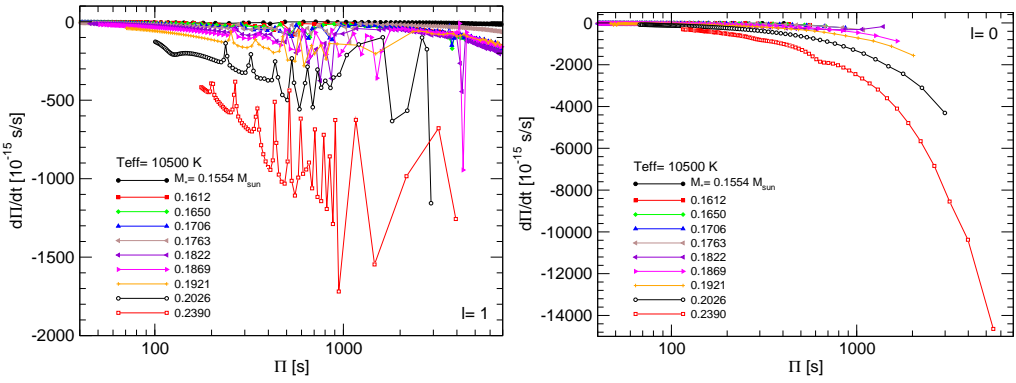

  \begin{center}
\subfigure{\includegraphics[width=.48\textwidth]{pre-ELMVs-dotp-p-10500.eps}}
\subfigure{\includegraphics[width=.48\textwidth]{pre-ELMVs-dotp-p-10500-radial.eps}}
  \end{center}
\caption{
  {\it Left:} The rate of period change versus the pulsation periods
  of $\ell= 1$ $g$, $p$, and $p-g$ mixed  modes, corresponding to
  low-mass pre-WD models characterized by an effective temperature of
  $T_{\rm eff} \sim 10500$ K and different stellar masses
  ($0.1554 \leq M_{\star}/M_{\sun} \leq 0.2390)$. {\it Right:} Same as in
  left panel, but for radial ($\ell= 0$) modes.}
\label{pre-ELMVs-dotp-p-10500}
\end{figure*}

The Brunt-V\"ais\"al\"a frequency in
the inner regions of our low-mass pre-WD models adopts very high values
\citep[see Figs. 2 and 3 of][]{2016A&A...588A..74C}.  As a consequence,
for these models there is a considerable range of
intermediate frequencies for which the modes behave like $g$ modes in
the inner parts of the star and  like $p$ modes in the outer parts.  A
similar situation is found in H-deficient pre-WD models
representative of  GW Vir stars before the evolutionary knee
\citep[][]{2006A&A...454..863C}. These  intermediate-frequency modes,
which are called ``$p-g$ mixed modes''
\citep{1974A&A....36..107S,1975PASJ...27..237O,1977A&A....58...41A},
are characterized by pulsation periods  between the short periods
of pure $p$ modes and the longer periods of $g$ modes.
Mixed modes and the associated phenomena  of ``avoided crossing'', in
which the periods approach each other quite closely without actually crossing
\citep{1977A&A....58...41A,1981MNRAS.194..229C}, have been extensively
investigated in the context of sub-giants and red giant pulsating
stars \citep[see, e.g.][]{2010Ap&SS.328...51C,2010Ap&SS.328..259D}. In
Fig. \ref{per-teff-01763} we show the pulsation periods of nonradial
$p$, $g$, and $p-g$ mixed modes with $\ell= 1$ as a function of
$T_{\rm eff}$  for He-core pre-WD models  with $M_{\star}= 0.1763
M_{\sun}$. The time evolution proceeds from low to high effective
temperatures, i.e., from right to left in the plot. Note that all the
pulsation periods decrease with increasing $T_{\rm eff}$, so the rates
of period change are negative for all the pulsation modes. According 
to Eq. (\ref{eq-dotp}), the second term must dominate over the first one
in order to $\dot{\Pi} < 0$. This reflects the fact that the models
are contracting as they evolve towards higher effective temperatures
(see Fig. \ref{HR-pre-ELMs}). The existence of multiples avoided
crossings is evident from Fig. \ref{per-teff-01763}
\footnote{In the case of radial modes, which is not shown, no avoided
  crossings are found.}.  The main effect of avoided crossing is that
during the approximation of the periods, they strongly change in
magnitude, so we would expect  $\dot{\Pi}$ to show
pronounced changes during an avoided crossing.  We display in
Fig. \ref{per-teff-01763-11000} the rates of period change as a function of
periods of $\ell= 1$ modes for a $0.1763 M_{\sun}$ pre-WD model at
$T_{\rm eff}= 11\,000$ K. The modes involved in avoided crossings are
emphasized with red dots. Clearly, some $p-g$ mixed modes that
take part of avoided 
crossings exhibit larger values (in modulus) than pure $g$ and $p$
modes that do not participate in avoided crossings.

\begin{figure*}[t]
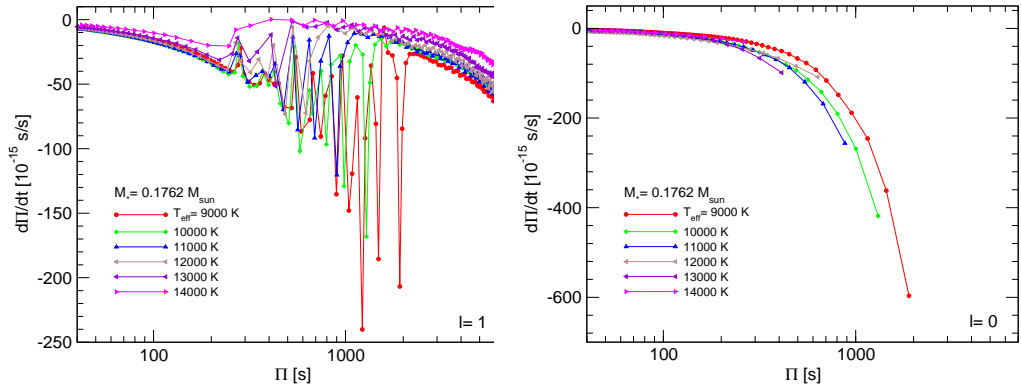

  \begin{center}
\subfigure{\includegraphics[width=.48\textwidth]{pre-ELMVs-dotp-p-01763.eps}}
\subfigure{\includegraphics[width=.48\textwidth]{pre-ELMVs-dotp-p-01763-radial.eps}}
  \end{center}
\caption{
{\it Left:}  The rate of period change versus the pulsation periods
  of $\ell= 1$ $g$, $p$, and $p-g$ mixed  modes, corresponding to
  low-mass pre-WD models characterized by a stellar
  mass $M_{\star}= 0.1703 M_{\sun}$
  and different effective temperatures. {\it Right:} Similar to
  left panel, but for radial ($\ell= 0$) modes.}
\label{pre-ELMVs-dotp-p-01763}
\end{figure*}

We now briefly explore the dependence of the $\dot{\Pi}$ values of our
low-mass pre-WD models with the stellar mass and the effective
temperature. As we mentioned, for all of our pre-WD models, the
complete set of radial and nonradial pulsation modes considered is
characterized by negative values of the rates of period change. In
Fig. \ref{pre-ELMVs-dotp-p-10500} we depict the rates of period change
as a function of the pulsation periods of $\ell= 1$ $g$, $p$, and
$p-g$ mixed  modes (left panel), corresponding to low-mass pre-WD
models with $T_{\rm eff} \sim 10500$ K and $0.1554 \leq
M_{\star}/M_{\sun} \leq 0.2390)$. The right panel shows the results
for radial modes. Our computations indicate that the magnitude of the
rates of period change is larger for higher stellar
masses. Specifically, $|\dot{\Pi}|$ is as large as $\sim 1.75 \times
10^{-12}$ s/s for the model with $M_{\star}= 0.2390 M_{\sun}$. In
comparison with the maximum values of $\dot{\Pi}$ predicted for ELMV
stars (Fig. \ref{max_rate_g_p_r} and Table \ref{table1}), the
magnitude of the rates of period change for nonradial  modes in the
case of pre-ELMV stars is roughly 20 times larger. In the case of
radial modes, on the other hand, the rates of period change for
pre-ELMVs   (right panel of  Fig. \ref{pre-ELMVs-dotp-p-10500}) are about
$3-4$ orders of magnitude larger than for ELMVs (right panel of Fig.
\ref{max_rate_g_p_r}).

\begin{figure*}[t]
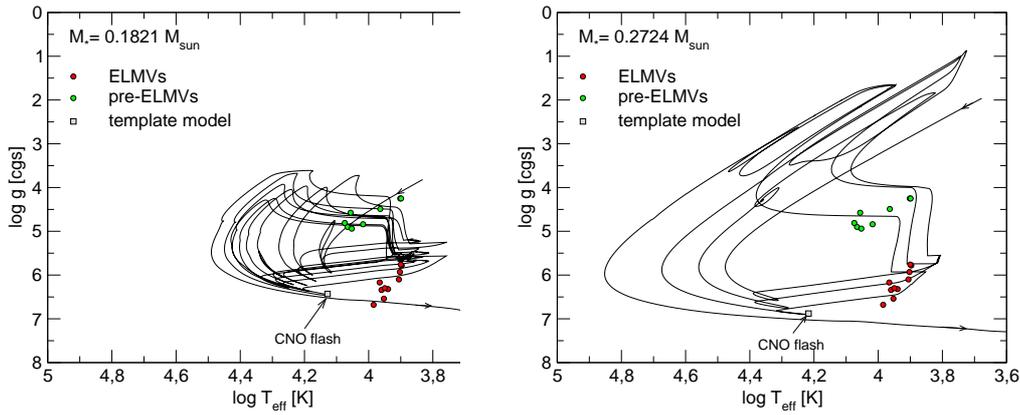

  \begin{center}
\subfigure{\includegraphics[width=.48\textwidth]{flash-018.eps}}
\subfigure{\includegraphics[width=.48\textwidth]{flash-027.eps}}
  \end{center}
\caption{
  {\it Left:} Evolutionary track of the $0.1821 M_{\sun}$ sequence
  in the $\log T_{\rm eff}- \log g$ diagram. Red (green) circles mark the
  location of the known ELMV (pre-ELMV) stars. The gray square 
  corresponds to a template model located shortly before the last
  CNO flash occurs. The arrows on the track indicate the sense of time
  evolution.
{\it Right:} Same as in left panel, but for the   $0.2724 M_{\sun}$ sequence.
}
\label{flashes}
\end{figure*}
 
In Fig. \ref{pre-ELMVs-dotp-p-01763} we show the $\dot{\Pi}$ values in
terms of $\Pi$ of $\ell= 1$ $g$, $p$, and $p-g$ mixed  modes (left
panel), corresponding to low-mass pre-WD models characterized by a
stellar mass $M_{\star}= 0.1703 M_{\sun}$ and different values of the
effective temperature ($8000 \lesssim T_{\rm eff} \lesssim 14000$
K). The right panel corresponds to radial modes. The magnitude of the
rates of period change for nonradial modes is larger for lower
effective temperatures. Interestingly enough, the reverse situation is
found in the case of radial modes. 

We conclude this Section by noting that, in general, the rates of
change of periods (both  of radial and nonradial modes) expected for
pre-ELMV stars are considerably larger ($\approx 1000-10000$ times)
than for ELMV stars. In practise, this indicates that it would be more
likely to achieve the measurement of $\dot{\Pi}$ for pre-ELMVs than
for ELMVs.

\subsection{Pre-CNO flash stages ($M_{\star} \gtrsim 0.18 M_{\sun}$)}
\label{CNO}

\begin{figure*}[t]
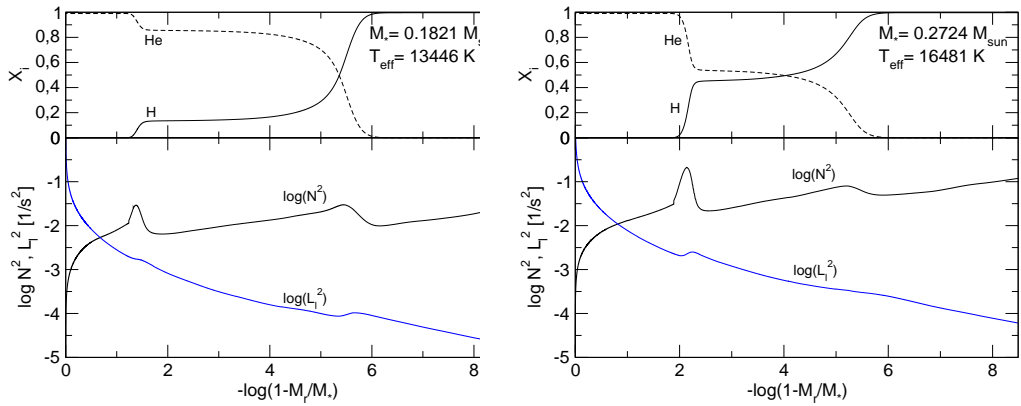

  \begin{center}
\subfigure{\includegraphics[width=.48\textwidth]{x-bv-018.eps}}
\subfigure{\includegraphics[width=.48\textwidth]{x-bv-027.eps}}
  \end{center}
\caption{
  {\it Left:} The internal chemical abundances of H and He (upper panel) and
  the logarithm of the squared Brunt-V\"ais\"al\"a and Lamb frequencies
  (lower panel)
  in terms of the mass coordinate [$-\log(1-M_r/M_{\star})$] corresponding to
  the pre-CNO flash template model with $M_{\star}= 0.1821 M_{\sun}$ and $T_{\rm eff}= 13446$ K
  whose location in the $\log T_{\rm eff}- \log g$ diagram is
  indicated in the left panel of Fig. \ref{flashes}.
  {\it Right:} Same as in left panel, but for the pre-CNO flash template
  model with $0.2724 M_{\sun}$ and $T_{\rm eff}= 16481$ K. Its
  location in the $\log T_{\rm eff}- \log g$ plane plane is
  indicated in the right panel of Fig. \ref{flashes}.
}
\label{x-bv}
\end{figure*}

Having explored in detail the theoretically expected properties of $\dot{\Pi}$
for the ELMV and pre-ELMV pulsating stars, here we focus on the
expected rates of period change for stellar models at
stages previous to the CNO flashes on the early cooling branches
(pre-flash stages). As it is shown in \citet{2013A&A...557A..19A}
(see their Fig. 4), the stages previous to the
occurrence of CNO flashes are characterized by a relatively slow evolution,
implying that there are chances to observe a star going through that
evolutionary phases. Put in numbers, the time spent by the star to evolve
between the maximum $T_{\rm eff}$ and the end of the last CNO flash in the case
of the  $0.1821 M_{\sun}$ sequence \citep[that is, the coloured portion
  of the track
in the lower panel of Fig. 4 of][]{2013A&A...557A..19A} is of
about $8 \times 10^{7}$ yrs. We emphasize that no pulsating low-mass WD star
is known until now that,
according to its $T_{\rm eff}$ and $\log g$ values, could be associated
with a pre-CNO flash stage. Even so, we believe that it is worth exploring
what the upper limit of the rates of period change for a star in
that evolutionary stage should be. In the computations of
\citet{2013A&A...557A..19A}, only model sequences with masses in the
range $0.186-0.362M_{\sun}$ experience CNO  flashes.
So, we restrict our analysis to models in that range of masses.
In Fig. \ref{flashes} we show 
the evolutionary tracks for the sequence with $0.1821 M_{\sun}$
(left) and $0.2724 M_{\sun}$ (right)
in the $\log T_{\rm eff}- \log g$ diagram. The gray squares 
correspond to representative models located shortly before the last
CNO flash for these sequences occurs. The chemical profiles
and the propagation diagram (the run of the logarithm of
the Brunt-V\"ais\"al\"a and Lamb frequencies in terms of the
mass coordinate) for these template models
are displayed in Fig. \ref{x-bv}. At variance with the case of
stellar models located at the stage of the pre-ELMVs (Sect. \ref{pre-wd})
in this case the propagation regions of $p$ and $g$ modes
are well delimited, so that there are no $p-g$  mixed modes, but only
pure $p$ and $g$ modes. 

\begin{figure*}[t]
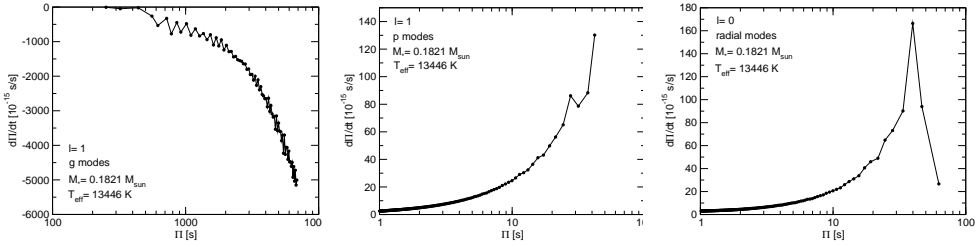

  \begin{center}
\subfigure{\includegraphics[width=.315\textwidth]{flashes-0182-gmodes.eps}}
\subfigure{\includegraphics[width=.3\textwidth]{flashes-0182-pmodes.eps}}
\subfigure{\includegraphics[width=.3\textwidth]{flashes-0182-rmodes.eps}}
  \end{center}
\caption{
  {\it Left:} The rate of period change (in units of $10^{-15}$ s/s) of
  $\ell= 1$  $g$ modes corresponding to the pre-CNO flash template model with
  $M_{\star}= 0.1821 M_{\sun}$ and $T_{\rm eff}= 13446$ K (see the left
  panels of Fig. \ref{x-bv}).
{\it Middle:} Same as in left panel, but for $\ell= 1$ $p$ modes.
{\it Right:} Same as in left panel, but for radial ($\ell= 0$) modes.
}
\label{flashes-0182}
\end{figure*}

\begin{figure*}[t]
  \begin{center}
\subfigure{\includegraphics[width=.315\textwidth]{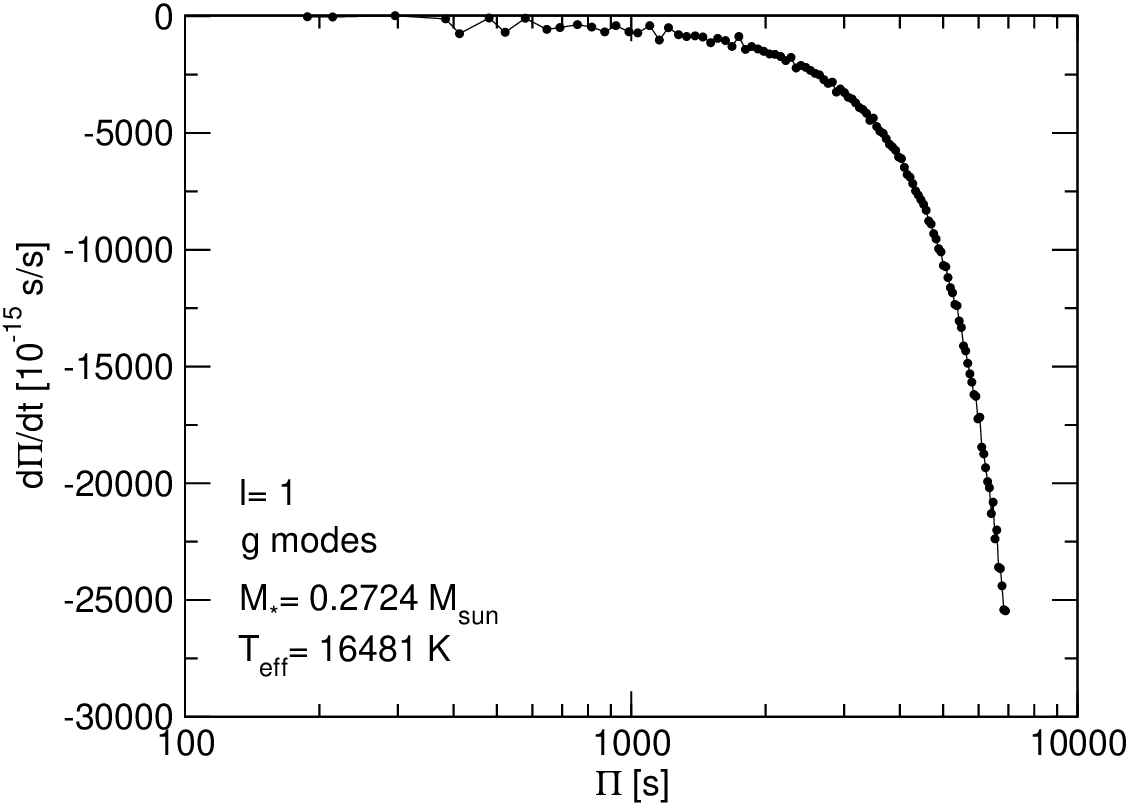}}
\subfigure{\includegraphics[width=.3\textwidth]{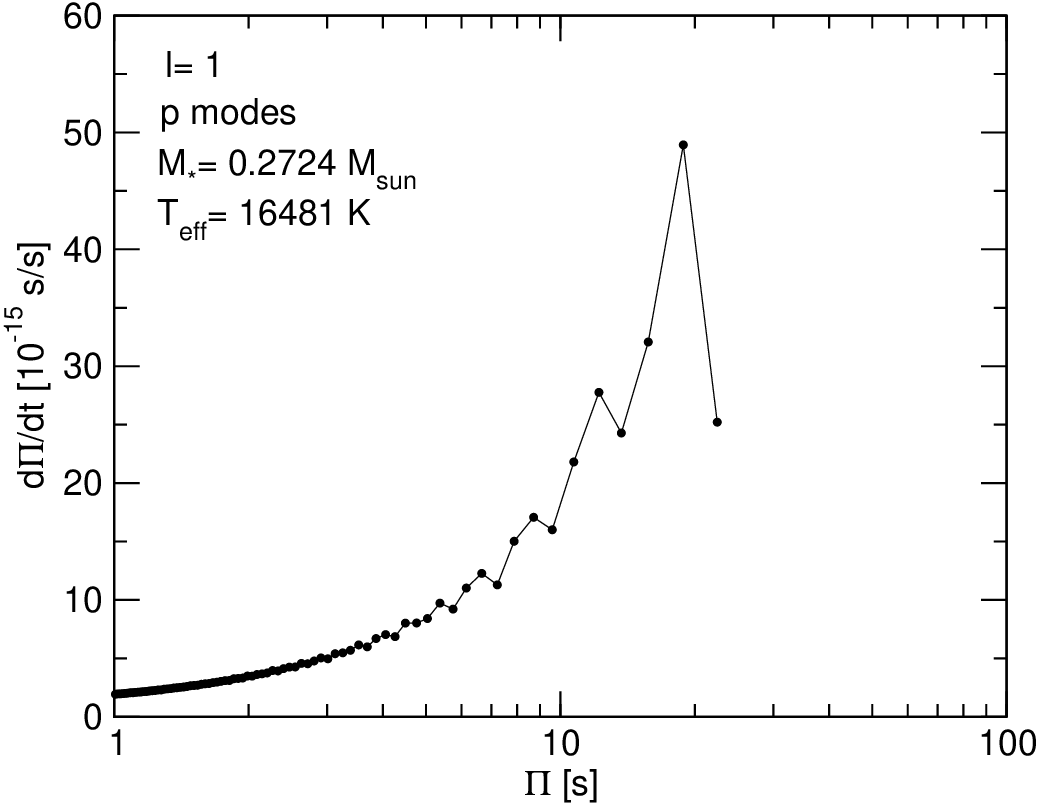}}
\subfigure{\includegraphics[width=.3\textwidth]{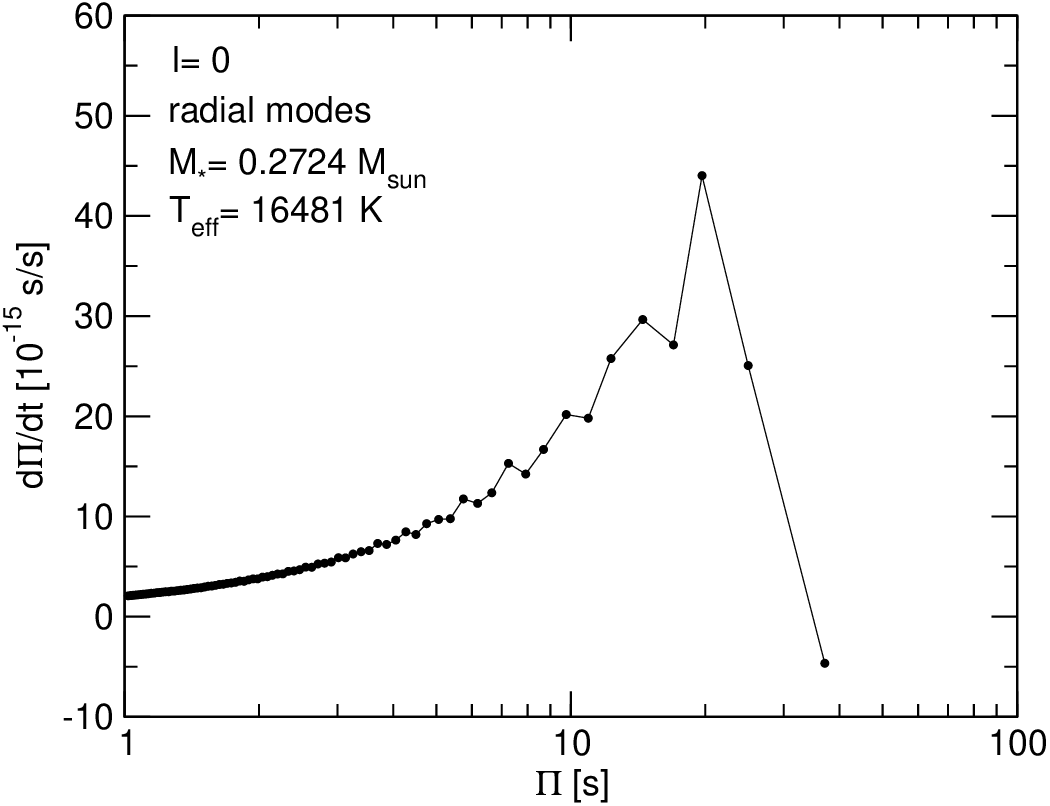}}
  \end{center}
\caption{
{\it Left:}  The rate of period change (in units of $10^{-15}$ s/s) of
  $\ell= 1$  $g$ modes corresponding to the pre-CNO flash template model with
  $M_{\star}= 0.2724 M_{\sun}$ and $T_{\rm eff}= 16481$ K (see the right
  panels of Fig. \ref{x-bv}). 
{\it Middle:} Same as in left panel, but for $\ell= 1$ $p$ modes.
{\it Right:} Same as in left panel, but for radial ($\ell= 0$) modes.
}
\label{flashes-0272}
\end{figure*}

Figs. \ref{flashes-0182} and \ref{flashes-0272} depict the rates of
period change in terms of periods corresponding to the template models
with ($M_{\star}= 0.1821 M_{\sun}, T_{\rm eff}= 13446$ K) and 
($M_{\star}= 0.2724 M_{\sun}, T_{\rm eff}= 16481$ K), respectively. Left panels
show the results for $g$ modes, while middle and right panels display
the results for $p$ and radial modes, respectively. The $\dot{\Pi}$ values
for $g$ modes are all negative, while for $p$ and radial modes  they
are all positive. For the three kinds of modes,
the magnitude of $\dot{\Pi}$ increases for increasing 
radial orders. Interestingly enough, the magnitude of $\dot{\Pi}$ for $g$ modes is
by far larger than for $p$ and radial modes. In particular, for the
range of periods considered in this work ($\Pi \lesssim 7000$ s),
we obtain  $|\dot{\Pi}|_{\rm max} \sim  5\times 10^{-12}$ s/s for $g$ modes
in the case of the $0.1821 M_{\sun}$ model,
and $|\dot{\Pi}|_{\rm max} \sim  2.5 \times 10^{-11}$ s/s
for the $0.2724 M_{\sun}$ model. In contrast, in the case of $p$ and
radial modes, $\dot{\Pi}$ adopts values of  $\sim  2 \times 10^{-13}$ s/s
at most. 

We conclude that, if a pulsating star  was evolving in stages prior to
a CNO flash, the measured rate of period change (if it could be
measured) would be of the order of $10^{-10}-10^{-11}$ s/s at
most. On the other hand, if the detected rate of period change   was
greater
than those values, the object would be evolving extremely fast through
any of the loops among flashes that are observed in Fig. \ref{flashes}
\citep[see also Fig. 2 of][]{2013A&A...557A..19A}. However, in that
case the evolution would be extremely fast, and therefore the
probability of catching such a star in that evolutionary stage
should be very low.

\section{Summary and conclusions}
\label{conclusions}

In this paper,  the fourth one of a series devoted to low-mass WD and
pre-WD stars, we presented for the first time a detailed assessment of
the theoretical temporal rates of period change of ELMV and pre-ELMV
pulsating stars.  Specifically, we have computed the rates of period
change for dipole ($\ell= 1$) nonradial $g$ and $p$ modes and also for
radial ($\ell= 0$) modes. We have considered periods up to $\sim 7000$
s, thus covering the period range detected in the known pulsating
low-mass WD stars.  We have considered evolutionary stages at the
final cooling branches (WD phase), evolutionary phases prior to the
occurrence of CNO flashes on the early cooling branches (pre-flash
stages), and the stages at constant luminosity following the end of
Roche lobe overflow (pre-WD phase).  The study is based on the modern
and physically sound evolutionary models of low-mass He-core WDs of
\citet{2013A&A...557A..19A}.

\begin{table*}
\centering
\caption{The absolute value of the \emph{maximum} expected
  rates of period change,
  $|\dot{\Pi}_{\rm max}|$ (s/s), and the sign, for
  nonradial $\ell= 1$ $g$ and $p$ modes (and $p-g$ mixed modes in the case of pre-ELMVs), and for
  radial  ($\ell= 0$) modes corresponding to low-mass WD and pre-WD models and
  also objects evolving just before the CNO flashes.}
\begin{tabular}{l|cc|cc|cc|cc}
\hline
\hline
\noalign{\smallskip}
Evolutionary phase & $|\dot{\Pi}_{\rm max}|$ ($g$ modes)   &   &  $|\dot{\Pi}_{\rm max}|$ ($p-g$ modes)    &   &  $|\dot{\Pi}_{\rm max}|$ ($p$ modes)    &   &
  $|\dot{\Pi}_{\rm max}|$ (radial modes) &  \\
\noalign{\smallskip}
\hline
\noalign{\smallskip}
pre-WD (pre-ELMVs) & $\sim 3 \times 10^{-13}$ & $(<0)$ & $\sim 5 \times 10^{-12}$ & $(<0)$ & $\sim 3 \times 10^{-12}$ & $(<0)$ & $\sim 5 \times 10^{-11}$ & $(<0)$\\
pre-CNO flashes & $\sim 3 \times 10^{-11}$ & $(<0)$ & $\cdots$ & &$\sim 2 \times 10^{-13}$ & $(>0)$&
$\sim 2 \times 10^{-13}$& $(>0)$\\
WD (ELMVs)      & $\sim 8 \times 10^{-14}$ & $(>0)$  & $\cdots$ & & $\sim 8 \times 10^{-15}$ & $(<0)$
& $\sim 3 \times 10^{-15}$ & $(<0)$\\
\noalign{\smallskip}
\hline
\hline
\end{tabular}
\label{table2}
\end{table*}

In Table \ref{table2} we summarize our
findings regarding the \emph{maximum}\footnote{That is, considering
  the maximum possible $|\dot{\Pi}|$ values not only for the specific
  models analyzed in
  previous sections, but globally for stellar models with all the
  masses and effective temperatures considered in this work.}
$\dot{\Pi}$ (in absolute value)
that we would expect for the different kinds of modes (nonradial $g$
and $p$ modes, and radial modes) for stars in the distinct evolutionary  
stages considered. We also include the sign of the rate of period
change in each case. We found that, in general, the theoretically expected
magnitude of the rates of period change of pure $g$ modes for pre-ELMVs
($\sim 3 \times 10^{-13}$) are larger  than for ELMVs ($\sim 8 \times 10^{-14}$).
Note that for pre-ELMVs, the largest rates of period change
($\sim 5 \times 10^{-12}$)
are associated to $p-g$ mixed modes involved in avoided crossings. 
In turn, the $\dot{\Pi}$ values of $g$ modes for models evolving shortly
before the occurrence of CNO flashes ($\sim 3 \times 10^{-11}$)
are by far larger than the maximum rates of period change
predicted for pure $g$ modes in pre-ELMV stars. 
If we focus on the ELMV models, we find that the rates of period change
for $M_{\star} \lesssim 0.18 M_{\odot}$ are 10 times smaller than for
more massive models. In the case of $p$
modes and radial modes, we found that the larger absolute values
correspond to pre-ELMV models, followed by the $\dot{\Pi}$ associated to
the pre-CNO flash phases. The smallest values of $\dot{\Pi}$ for these
kinds of modes correspond to ELMV models.

Based on the theoretically predicted values of $\dot{\Pi}$
presented in this work, we conclude that any future measurement of a rate of
period change for a given pulsating low-mass pre-WD or WD star could shed light
about its evolutionary status. In particular, it could be
possible to distinguish a star that is in its pre-WD phase,
if it is evolving in stages just prior to a H flash, or if it
is already settled on its final cooling stage as a WD. Eventually,
it would also be possible (although less likely) to
distinguish whether or not a star is in
its final cooling branch has a extremely low mass ($M_{\star} \lesssim
0.18 M_{\sun}$). Finally, a measured value of $\dot{\Pi}$ larger
than $\sim 10^{-10}$ s/s would imply that the star is still evolving
rapidly between the CNO flashes.

We close the paper by noting that the recent results
of \citet{2016arXiv161206390B} cast some doubts on the ELMV classification
of several stars in the ELM Survey, apart from just SDSS J135512+195645.
These authors find an
overabundance of pulsating stars with ELM-like spectra 
at $T_{\rm eff} \lesssim 9000$ K that lack radial-velocity variations,
and so, it cannot be confirmed that they are in binary systems, as
we would expect on the grounds of the currently accepted evolutionary
channels for these stars.  Therefore, in view of the systematic
difficulties in the spectroscopic classification, an eventual
measurement of $\dot{\Pi}$ could help to confirm that a given
pulsating star is a genuine ELM WD.

\begin{acknowledgements}
 We wish to thank our anonymous referee for his/her constructive
comments and suggestions that improved the original version of
the paper. We also thanks Keaton Bell and JJ Hermes for the exchange of emails
with valuable comments that enriched the content of this study.
Part of this work was supported by AGENCIA through the
Programa de Modernizaci\'on Tecnol\'ogica BID 1728/OC-AR, and by the
PIP 112-200801-00940 grant from CONICET. This research has made use of
NASA Astrophysics Data System.        
\end{acknowledgements}


\bibliographystyle{aa} 
\bibliography{paper} 

\begin{thebibliography}{99}
\expandafter\ifx\csname natexlab\endcsname\relax\def\natexlab#1{#1}\fi

\bibitem[{{Aizenman} {et~al.}(1977){Aizenman}, {Smeyers}, \&
  {Weigert}}]{1977A&A....58...41A}
{Aizenman}, M., {Smeyers}, P., \& {Weigert}, A. 1977, \aap, 58, 41

\bibitem[{{Althaus} {et~al.}(2015){Althaus}, {Camisassa}, {Miller Bertolami},
  {C{\'o}rsico}, \& {Garc{\'{\i}}a-Berro}}]{2015A&A...576A...9A}
{Althaus}, L.~G., {Camisassa}, M.~E., {Miller Bertolami}, M.~M., {C{\'o}rsico},
  A.~H., \& {Garc{\'{\i}}a-Berro}, E. 2015, \aap, 576, A9

\bibitem[{{Althaus} {et~al.}(2010){Althaus}, {C{\'o}rsico}, {Isern}, \&
  {Garc{\'{\i}}a-Berro}}]{review}
{Althaus}, L.~G., {C{\'o}rsico}, A.~H., {Isern}, J., \& {Garc{\'{\i}}a-Berro},
  E. 2010, \aapr, 18, 471

\bibitem[{{Althaus} {et~al.}(2008){Althaus}, {C{\'o}rsico}, {Miller Bertolami},
  {Garc{\'{\i}}a-Berro}, \& {Kepler}}]{2008ApJ...677L..35A}
{Althaus}, L.~G., {C{\'o}rsico}, A.~H., {Miller Bertolami}, M.~M.,
  {Garc{\'{\i}}a-Berro}, E., \& {Kepler}, S.~O. 2008, \apjl, 677, L35

\bibitem[{{Althaus} {et~al.}(2016){Althaus}, {De Ger{\'o}nimo}, {C{\'o}rsico},
  {Torres}, \& {Garc{\'{\i}}a--Berro}}]{2016arXiv161106191A}
{Althaus}, L.~G., {De Ger{\'o}nimo}, F.~C., {C{\'o}rsico}, A.~H., {Torres}, S.,
  \& {Garc{\'{\i}}a--Berro}, E. 2016, ArXiv e-prints

\bibitem[{{Althaus} {et~al.}(2013){Althaus}, {Miller Bertolami}, \&
  {C{\'o}rsico}}]{2013A&A...557A..19A}
{Althaus}, L.~G., {Miller Bertolami}, M.~M., \& {C{\'o}rsico}, A.~H. 2013,
  \aap, 557, A19

\bibitem[{{Althaus} {et~al.}(2009){Althaus}, {Panei}, {Romero}, {Rohrmann},
  {C{\'o}rsico}, {Garc{\'{\i}}a-Berro}, \& {Miller
  Bertolami}}]{2009A&A...502..207A}
{Althaus}, L.~G., {Panei}, J.~A., {Romero}, A.~D., {et~al.} 2009, \aap, 502,
  207

\bibitem[{{Althaus} {et~al.}(2001){Althaus}, {Serenelli}, \&
  {Benvenuto}}]{2001MNRAS.323..471A}
{Althaus}, L.~G., {Serenelli}, A.~M., \& {Benvenuto}, O.~G. 2001, \mnras, 323,
  471

\bibitem[{{Althaus} {et~al.}(2005){Althaus}, {Serenelli}, {Panei},
  {C{\'o}rsico}, {Garc{\'{\i}}a-Berro}, \&
  {Sc{\'o}ccola}}]{2005A&A...435..631A}
{Althaus}, L.~G., {Serenelli}, A.~M., {Panei}, J.~A., {et~al.} 2005, \aap, 435,
  631

\bibitem[{{Bassa} {et~al.}(2003){Bassa}, {van Kerkwijk}, \&
  {Kulkarni}}]{2003A&A...403.1067B}
{Bassa}, C.~G., {van Kerkwijk}, M.~H., \& {Kulkarni}, S.~R. 2003, \aap, 403,
  1067

\bibitem[{{Battich} {et~al.}(2016){Battich}, {C{\'o}rsico}, {Althaus}, \&
  {Miller Bertolami}}]{2016JCAP...08..062B}
{Battich}, T., {C{\'o}rsico}, A.~H., {Althaus}, L.~G., \& {Miller Bertolami},
  M.~M. 2016, \jcap, 8, 062

\bibitem[{{Bell} {et~al.}(2016){Bell}, {Gianninas}, {Hermes}, {Winget},
  {Kilic}, {Montgomery}, {Castanheira}, {Vanderbosch}, {Winget}, \&
  {Brown}}]{2016arXiv161206390B}
{Bell}, K.~J., {Gianninas}, A., {Hermes}, J.~J., {et~al.} 2016, ArXiv e-prints

\bibitem[{{Bell} {et~al.}(2015){Bell}, {Kepler}, {Montgomery}, {Hermes},
  {Harrold}, \& {Winget}}]{2015ASPC..493..217B}
{Bell}, K.~J., {Kepler}, S.~O., {Montgomery}, M.~H., {et~al.} 2015, in
  Astronomical Society of the Pacific Conference Series, Vol. 493, 19th
  European Workshop on White Dwarfs, ed. P.~{Dufour}, P.~{Bergeron}, \&
  G.~{Fontaine}, 217

\bibitem[{{Bischoff-Kim} {et~al.}(2008){Bischoff-Kim}, {Montgomery}, \&
  {Winget}}]{2008ApJ...675.1512B}
{Bischoff-Kim}, A., {Montgomery}, M.~H., \& {Winget}, D.~E. 2008, \apj, 675,
  1512

\bibitem[{{Bradley}(1996)}]{B96}
{Bradley}, P.~A. 1996, \apj, 468, 350

\bibitem[{{Bradley} {et~al.}(1992){Bradley}, {Winget}, \&
  {Wood}}]{1992ApJ...391L..33B}
{Bradley}, P.~A., {Winget}, D.~E., \& {Wood}, M.~A. 1992, \apjl, 391, L33

\bibitem[{{Brassard} {et~al.}(1991){Brassard}, {Fontaine}, {Wesemael},
  {Kawaler}, \& {Tassoul}}]{1991ApJ...367..601B}
{Brassard}, P., {Fontaine}, G., {Wesemael}, F., {Kawaler}, S.~D., \& {Tassoul},
  M. 1991, \apj, 367, 601

\bibitem[{{Breton} {et~al.}(2012){Breton}, {Rappaport}, {van Kerkwijk}, \&
  {Carter}}]{2012ApJ...748..115B}
{Breton}, R.~P., {Rappaport}, S.~A., {van Kerkwijk}, M.~H., \& {Carter}, J.~A.
  2012, \apj, 748, 115

\bibitem[{{Brickhill}(1991)}]{1991MNRAS.251..673B}
{Brickhill}, A.~J. 1991, \mnras, 251, 673

\bibitem[{{Brown} {et~al.}(2016{\natexlab{a}}){Brown}, {Gianninas}, {Kilic},
  {Kenyon}, \& {Allende Prieto}}]{2016ApJ...818..155B}
{Brown}, W.~R., {Gianninas}, A., {Kilic}, M., {Kenyon}, S.~J., \& {Allende
  Prieto}, C. 2016{\natexlab{a}}, \apj, 818, 155

\bibitem[{{Brown} {et~al.}(2013){Brown}, {Kilic}, {Allende Prieto},
  {Gianninas}, \& {Kenyon}}]{2013ApJ...769...66B}
{Brown}, W.~R., {Kilic}, M., {Allende Prieto}, C., {Gianninas}, A., \&
  {Kenyon}, S.~J. 2013, \apj, 769, 66

\bibitem[{{Brown} {et~al.}(2010){Brown}, {Kilic}, {Allende Prieto}, \&
  {Kenyon}}]{2010ApJ...723.1072B}
{Brown}, W.~R., {Kilic}, M., {Allende Prieto}, C., \& {Kenyon}, S.~J. 2010,
  \apj, 723, 1072

\bibitem[{{Brown} {et~al.}(2012){Brown}, {Kilic}, {Allende Prieto}, \&
  {Kenyon}}]{2012ApJ...744..142B}
---. 2012, \apj, 744, 142

\bibitem[{{Brown} {et~al.}(2016{\natexlab{b}}){Brown}, {Kilic}, {Kenyon}, \&
  {Gianninas}}]{2016ApJ...824...46B}
{Brown}, W.~R., {Kilic}, M., {Kenyon}, S.~J., \& {Gianninas}, A.
  2016{\natexlab{b}}, \apj, 824, 46

\bibitem[{{Burgers}(1969)}]{1969fecg.book.....B}
{Burgers}, J.~M. 1969, {Flow Equations for Composite Gases} (New York: Academic
  Press)

\bibitem[{{Carter} {et~al.}(2011){Carter}, {Rappaport}, \&
  {Fabrycky}}]{2011ApJ...728..139C}
{Carter}, J.~A., {Rappaport}, S., \& {Fabrycky}, D. 2011, \apj, 728, 139

\bibitem[{{Cassisi} {et~al.}(2007){Cassisi}, {Potekhin}, {Pietrinferni},
  {Catelan}, \& {Salaris}}]{2007ApJ...661.1094C}
{Cassisi}, S., {Potekhin}, A.~Y., {Pietrinferni}, A., {Catelan}, M., \&
  {Salaris}, M. 2007, \apj, 661, 1094

\bibitem[{{Christensen-Dalsgaard}(1981)}]{1981MNRAS.194..229C}
{Christensen-Dalsgaard}, J. 1981, \mnras, 194, 229

\bibitem[{{Christensen-Dalsgaard} \& {Houdek}(2010)}]{2010Ap&SS.328...51C}
{Christensen-Dalsgaard}, J. \& {Houdek}, G. 2010, \apss, 328, 51

\bibitem[{{C{\'o}rsico} \& {Althaus}(2004)}]{2004A&A...428..159C}
{C{\'o}rsico}, A.~H. \& {Althaus}, L.~G. 2004, \aap, 428, 159

\bibitem[{{C{\'o}rsico} \& {Althaus}(2006)}]{2006A&A...454..863C}
---. 2006, \aap, 454, 863

\bibitem[{{C{\'o}rsico} \& {Althaus}(2014{\natexlab{a}})}]{2014A&A...569A.106C}
---. 2014{\natexlab{a}}, \aap, 569, A106

\bibitem[{{C{\'o}rsico} \& {Althaus}(2014{\natexlab{b}})}]{2014ApJ...793L..17C}
---. 2014{\natexlab{b}}, \apjl, 793, L17

\bibitem[{{C{\'o}rsico} \& {Althaus}(2016)}]{2016A&A...585A...1C}
---. 2016, \aap, 585, A1

\bibitem[{{C{\'o}rsico} {et~al.}(2013){C{\'o}rsico}, {Althaus},
  {Garc{\'{\i}}a-Berro}, \& {Romero}}]{2013JCAP...06..032C}
{C{\'o}rsico}, A.~H., {Althaus}, L.~G., {Garc{\'{\i}}a-Berro}, E., \& {Romero},
  A.~D. 2013, \jcap, 6, 032

\bibitem[{{C{\'o}rsico} {et~al.}(2008){C{\'o}rsico}, {Althaus}, {Kepler},
  {Costa}, \& {Miller Bertolami}}]{2008A&A...478..869C}
{C{\'o}rsico}, A.~H., {Althaus}, L.~G., {Kepler}, S.~O., {Costa}, J.~E.~S., \&
  {Miller Bertolami}, M.~M. 2008, \aap, 478, 869

\bibitem[{{C{\'o}rsico} {et~al.}(2014){C{\'o}rsico}, {Althaus}, {Miller
  Bertolami}, {Kepler}, \& {Garc{\'{\i}}a-Berro}}]{2014JCAP...08..054C}
{C{\'o}rsico}, A.~H., {Althaus}, L.~G., {Miller Bertolami}, M.~M., {Kepler},
  S.~O., \& {Garc{\'{\i}}a-Berro}, E. 2014, \jcap, 8, 054

\bibitem[{{C{\'o}rsico} {et~al.}(2012{\natexlab{a}}){C{\'o}rsico}, {Althaus},
  {Miller Bertolami}, {Romero}, {Garc{\'{\i}}a-Berro}, {Isern}, \&
  {Kepler}}]{2012MNRAS.424.2792C}
{C{\'o}rsico}, A.~H., {Althaus}, L.~G., {Miller Bertolami}, M.~M., {et~al.}
  2012{\natexlab{a}}, \mnras, 424, 2792

\bibitem[{{C{\'o}rsico} {et~al.}(2012{\natexlab{b}}){C{\'o}rsico}, {Althaus},
  {Romero}, {Mukadam}, {Garc{\'{\i}}a-Berro}, {Isern}, {Kepler}, \&
  {Corti}}]{2012JCAP...12..010C}
{C{\'o}rsico}, A.~H., {Althaus}, L.~G., {Romero}, A.~D., {et~al.}
  2012{\natexlab{b}}, \jcap, 12, 10

\bibitem[{{C{\'o}rsico} {et~al.}(2016{\natexlab{a}}){C{\'o}rsico}, {Althaus},
  {Serenelli}, {Kepler}, {Jeffery}, \& {Corti}}]{2016A&A...588A..74C}
{C{\'o}rsico}, A.~H., {Althaus}, L.~G., {Serenelli}, A.~M., {et~al.}
  2016{\natexlab{a}}, \aap, 588, A74

\bibitem[{{C{\'o}rsico} {et~al.}(2001){C{\'o}rsico}, {Benvenuto}, {Althaus},
  {Isern}, \& {Garc{\'{\i}}a-Berro}}]{2001NewA....6..197C}
{C{\'o}rsico}, A.~H., {Benvenuto}, O.~G., {Althaus}, L.~G., {Isern}, J., \&
  {Garc{\'{\i}}a-Berro}, E. 2001, \na, 6, 197

\bibitem[{{C{\'o}rsico} {et~al.}(2007){C{\'o}rsico}, {Miller Bertolami},
  {Althaus}, {Vauclair}, \& {Werner}}]{2007A&A...475..619C}
{C{\'o}rsico}, A.~H., {Miller Bertolami}, M.~M., {Althaus}, L.~G., {Vauclair},
  G., \& {Werner}, K. 2007, \aap, 475, 619

\bibitem[{{C{\'o}rsico} {et~al.}(2016{\natexlab{b}}){C{\'o}rsico}, {Romero},
  {Althaus}, {Garc{\'{\i}}a-Berro}, {Isern}, {Kepler}, {Miller Bertolami},
  {Sullivan}, \& {Chote}}]{2016JCAP...07..036C}
{C{\'o}rsico}, A.~H., {Romero}, A.~D., {Althaus}, L.~G., {et~al.}
  2016{\natexlab{b}}, \jcap, 7, 036

\bibitem[{{C{\'o}rsico} {et~al.}(2012{\natexlab{c}}){C{\'o}rsico}, {Romero},
  {Althaus}, \& {Hermes}}]{2012A&A...547A..96C}
{C{\'o}rsico}, A.~H., {Romero}, A.~D., {Althaus}, L.~G., \& {Hermes}, J.~J.
  2012{\natexlab{c}}, \aap, 547, A96

\bibitem[{{Corti} {et~al.}(2016){Corti}, {Kanaan}, {C{\'o}rsico}, {Kepler},
  {Althaus}, {Koester}, \& {S{\'a}nchez Arias}}]{2016A&A...587L...5C}
{Corti}, M.~A., {Kanaan}, A., {C{\'o}rsico}, A.~H., {et~al.} 2016, \aap, 587,
  L5

\bibitem[{{Costa} \& {Kepler}(2008)}]{2008A&A...489.1225C}
{Costa}, J.~E.~S. \& {Kepler}, S.~O. 2008, \aap, 489, 1225

\bibitem[{{Deheuvels} \& {Michel}(2010)}]{2010Ap&SS.328..259D}
{Deheuvels}, S. \& {Michel}, E. 2010, \apss, 328, 259

\bibitem[{{Dziembowski}(1971)}]{1971AcA....21..289D}
{Dziembowski}, W.~A. 1971, \actaa, 21, 289

\bibitem[{{Fontaine} \& {Brassard}(2008)}]{FB08}
{Fontaine}, G. \& {Brassard}, P. 2008, \pasp, 120, 1043

\bibitem[{{Garc{\'{\i}}a-Berro} {et~al.}(2010){Garc{\'{\i}}a-Berro}, {Torres},
  {Althaus}, {Renedo}, {Lor{\'e}n-Aguilar}, {C{\'o}rsico}, {Rohrmann},
  {Salaris}, \& {Isern}}]{2010Natur.465..194G}
{Garc{\'{\i}}a-Berro}, E., {Torres}, S., {Althaus}, L.~G., {et~al.} 2010, \nat,
  465, 194

\bibitem[{{Gianninas} {et~al.}(2016){Gianninas}, {Curd}, {Fontaine}, {Brown},
  \& {Kilic}}]{2016ApJ...822L..27G}
{Gianninas}, A., {Curd}, B., {Fontaine}, G., {Brown}, W.~R., \& {Kilic}, M.
  2016, \apjl, 822, L27

\bibitem[{{Gianninas} {et~al.}(2014){Gianninas}, {Dufour}, {Kilic}, {Brown},
  {Bergeron}, \& {Hermes}}]{2014ApJ...794...35G}
{Gianninas}, A., {Dufour}, P., {Kilic}, M., {et~al.} 2014, \apj, 794, 35

\bibitem[{{Gianninas} {et~al.}(2015){Gianninas}, {Kilic}, {Brown}, {Canton}, \&
  {Kenyon}}]{2015ApJ...812..167G}
{Gianninas}, A., {Kilic}, M., {Brown}, W.~R., {Canton}, P., \& {Kenyon}, S.~J.
  2015, \apj, 812, 167

\bibitem[{{Haft} {et~al.}(1994){Haft}, {Raffelt}, \&
  {Weiss}}]{1994ApJ...425..222H}
{Haft}, M., {Raffelt}, G., \& {Weiss}, A. 1994, \apj, 425, 222

\bibitem[{{Hansen} {et~al.}(2013){Hansen}, {Kalirai}, {Anderson}, {Dotter},
  {Richer}, {Rich}, {Shara}, {Fahlman}, {Hurley}, {King}, {Reitzel}, \&
  {Stetson}}]{2013Natur.500...51H}
{Hansen}, B.~M.~S., {Kalirai}, J.~S., {Anderson}, J., {et~al.} 2013, \nat, 500,
  51

\bibitem[{{Harris} {et~al.}(2006){Harris}, {Munn}, {Kilic}, {Liebert},
  {Williams}, {von Hippel}, {Levine}, {Monet}, {Eisenstein}, {Kleinman},
  {Metcalfe}, {Nitta}, {Winget}, {Brinkmann}, {Fukugita}, {Knapp}, {Lupton},
  {Smith}, \& {Schneider}}]{2006AJ....131..571H}
{Harris}, H.~C., {Munn}, J.~A., {Kilic}, M., {et~al.} 2006, \aj, 131, 571

\bibitem[{{Hermes} {et~al.}(2013{\natexlab{a}}){Hermes}, {Montgomery},
  {Gianninas}, {Winget}, {Brown}, {Harrold}, {Bell}, {Kenyon}, {Kilic}, \&
  {Castanheira}}]{2013MNRAS.436.3573H}
{Hermes}, J.~J., {Montgomery}, M.~H., {Gianninas}, A., {et~al.}
  2013{\natexlab{a}}, \mnras, 436, 3573

\bibitem[{{Hermes} {et~al.}(2013{\natexlab{b}}){Hermes}, {Montgomery},
  {Mullally}, {Winget}, \& {Bischoff-Kim}}]{2013ApJ...766...42H}
{Hermes}, J.~J., {Montgomery}, M.~H., {Mullally}, F., {Winget}, D.~E., \&
  {Bischoff-Kim}, A. 2013{\natexlab{b}}, \apj, 766, 42

\bibitem[{{Hermes} {et~al.}(2013{\natexlab{c}}){Hermes}, {Montgomery},
  {Winget}, {Brown}, {Gianninas}, {Kilic}, {Kenyon}, {Bell}, \&
  {Harrold}}]{2013ApJ...765..102H}
{Hermes}, J.~J., {Montgomery}, M.~H., {Winget}, D.~E., {et~al.}
  2013{\natexlab{c}}, \apj, 765, 102

\bibitem[{{Hermes} {et~al.}(2012){Hermes}, {Montgomery}, {Winget}, {Brown},
  {Kilic}, \& {Kenyon}}]{2012ApJ...750L..28H}
---. 2012, \apjl, 750, L28

\bibitem[{{Iglesias} \& {Rogers}(1996)}]{1996ApJ...464..943I}
{Iglesias}, C.~A. \& {Rogers}, F.~J. 1996, \apj, 464, 943

\bibitem[{{Isern} {et~al.}(1998){Isern}, {Garc{\'{\i}}a-Berro}, {Hernanz},
  {Mochkovitch}, \& {Torres}}]{1998ApJ...503..239I}
{Isern}, J., {Garc{\'{\i}}a-Berro}, E., {Hernanz}, M., {Mochkovitch}, R., \&
  {Torres}, S. 1998, \apj, 503, 239

\bibitem[{{Isern} {et~al.}(1992){Isern}, {Hernanz}, \&
  {Garc{\'{\i}}a-Berro}}]{Isern92}
{Isern}, J., {Hernanz}, M., \& {Garc{\'{\i}}a-Berro}, E. 1992, \apjl, 392, L23

\bibitem[{{Istrate} {et~al.}(2016{\natexlab{a}}){Istrate}, {Fontaine},
  {Gianninas}, {Grassitelli}, {Marchant}, {Tauris}, \&
  {Langer}}]{2016A&A...595L..12I}
{Istrate}, A.~G., {Fontaine}, G., {Gianninas}, A., {et~al.} 2016{\natexlab{a}},
  \aap, 595, L12

\bibitem[{{Istrate} {et~al.}(2016{\natexlab{b}}){Istrate}, {Marchant},
  {Tauris}, {Langer}, {Stancliffe}, \& {Grassitelli}}]{2016A&A...595A..35I}
{Istrate}, A.~G., {Marchant}, P., {Tauris}, T.~M., {et~al.} 2016{\natexlab{b}},
  \aap, 595, A35

\bibitem[{{Itoh} {et~al.}(1996){Itoh}, {Hayashi}, {Nishikawa}, \&
  {Kohyama}}]{1996ApJS..102..411I}
{Itoh}, N., {Hayashi}, H., {Nishikawa}, A., \& {Kohyama}, Y. 1996, \apjs, 102,
  411

\bibitem[{{Jeffery} \& {Saio}(2013)}]{2013MNRAS.435..885J}
{Jeffery}, C.~S. \& {Saio}, H. 2013, \mnras, 435, 885

\bibitem[{{Kawaler} \& {Bradley}(1994)}]{1994ApJ...427..415K}
{Kawaler}, S.~D. \& {Bradley}, P.~A. 1994, \apj, 427, 415

\bibitem[{{Kawaler} {et~al.}(1986){Kawaler}, {Winget}, {Iben}, \&
  {Hansen}}]{1986ApJ...302..530K}
{Kawaler}, S.~D., {Winget}, D.~E., {Iben}, Jr., I., \& {Hansen}, C.~J. 1986,
  \apj, 302, 530

\bibitem[{{Kepler}(2012)}]{Kea12}
{Kepler}, S.~O. 2012, in Astronomical Society of the Pacific Conference Series,
  Vol. 462, Progress in Solar/Stellar Physics with Helio- and Asteroseismology,
  ed. H.~{Shibahashi}, 322

\bibitem[{{Kepler} {et~al.}(2005){Kepler}, {Costa}, {Castanheira}, {Winget},
  {Mullally}, {Nather}, {Kilic}, {von Hippel}, {Mukadam}, \&
  {Sullivan}}]{2005ApJ...634.1311K}
{Kepler}, S.~O., {Costa}, J.~E.~S., {Castanheira}, B.~G., {et~al.} 2005, \apj,
  634, 1311

\bibitem[{{Kilic} {et~al.}(2011){Kilic}, {Brown}, {Allende Prieto},
  {Ag{\"u}eros}, {Heinke}, \& {Kenyon}}]{2011ApJ...727....3K}
{Kilic}, M., {Brown}, W.~R., {Allende Prieto}, C., {et~al.} 2011, \apj, 727, 3

\bibitem[{{Kilic} {et~al.}(2012){Kilic}, {Brown}, {Allende Prieto}, {Kenyon},
  {Heinke}, {Ag{\"u}eros}, \& {Kleinman}}]{2012ApJ...751..141K}
---. 2012, \apj, 751, 141

\bibitem[{{Kilic} {et~al.}(2015){Kilic}, {Hermes}, {Gianninas}, \&
  {Brown}}]{2015MNRAS.446L..26K}
{Kilic}, M., {Hermes}, J.~J., {Gianninas}, A., \& {Brown}, W.~R. 2015, \mnras,
  446, L26

\bibitem[{{Koester} {et~al.}(2009){Koester}, {Voss}, {Napiwotzki},
  {Christlieb}, {Homeier}, {Lisker}, {Reimers}, \&
  {Heber}}]{2009A&A...505..441K}
{Koester}, D., {Voss}, B., {Napiwotzki}, R., {et~al.} 2009, \aap, 505, 441

\bibitem[{{Magni} \& {Mazzitelli}(1979)}]{1979A&A....72..134M}
{Magni}, G. \& {Mazzitelli}, I. 1979, \aap, 72, 134

\bibitem[{{Maxted} {et~al.}(2011){Maxted}, {Anderson}, {Burleigh}, {Collier
  Cameron}, {Heber}, {G{\"a}nsicke}, {Geier}, {Kupfer}, {Marsh}, {Nelemans},
  {O'Toole}, {{\O}stensen}, {Smalley}, \& {West}}]{2011MNRAS.418.1156M}
{Maxted}, P.~F.~L., {Anderson}, D.~R., {Burleigh}, M.~R., {et~al.} 2011,
  \mnras, 418, 1156

\bibitem[{{Maxted} {et~al.}(2014){Maxted}, {Serenelli}, {Marsh}, {Catal{\'a}n},
  {Mahtani}, \& {Dhillon}}]{2014MNRAS.444..208M}
{Maxted}, P.~F.~L., {Serenelli}, A.~M., {Marsh}, T.~R., {et~al.} 2014, \mnras,
  444, 208

\bibitem[{{Maxted} {et~al.}(2013){Maxted}, {Serenelli}, {Miglio}, {Marsh},
  {Heber}, {Dhillon}, {Littlefair}, {Copperwheat}, {Smalley}, {Breedt}, \&
  {Schaffenroth}}]{2013Natur.498..463M}
{Maxted}, P.~F.~L., {Serenelli}, A.~M., {Miglio}, A., {et~al.} 2013, \nat, 498,
  463

\bibitem[{{Mestel}(1952)}]{1952MNRAS.112..583M}
{Mestel}, L. 1952, \mnras, 112, 583

\bibitem[{{Mukadam} {et~al.}(2013){Mukadam}, {Bischoff-Kim}, {Fraser},
  {C{\'o}rsico}, {Montgomery}, {Kepler}, {Romero}, {Winget}, {Hermes},
  {Riecken}, {Kronberg}, {Winget}, {Falcon}, {Chandler}, {Kuehne}, {Sullivan},
  {Reaves}, {von Hippel}, {Mullally}, {Shipman}, {Thompson}, {Silvestri}, \&
  {Hynes}}]{2013ApJ...771...17M}
{Mukadam}, A.~S., {Bischoff-Kim}, A., {Fraser}, O., {et~al.} 2013, \apj, 771,
  17

\bibitem[{{Mukadam} {et~al.}(2003){Mukadam}, {Kepler}, {Winget}, {Nather},
  {Kilic}, {Mullally}, {von Hippel}, {Kleinman}, {Nitta}, {Guzik}, {Bradley},
  {Matthews}, {Sekiguchi}, {Sullivan}, {Sullivan}, {Shobbrook}, {Birch},
  {Jiang}, {Xu}, {Joshi}, {Ashoka}, {Ibbetson}, {Leibowitz}, {Ofek}, {Mei{\v
  s}tas}, {Janulis}, {Ali{\v s}auskas}, {Kalytis}, {Handler}, {Kilkenny},
  {O'Donoghue}, {Kurtz}, {M{\"u}ller}, {Moskalik}, {Og{\l}oza}, {Zo{\l}a},
  {Krzesi{\'n}ski}, {Johannessen}, {Gonzalez-Perez}, {Solheim}, {Silvotti},
  {Bernabei}, {Vauclair}, {Dolez}, {Fu}, {Chevreton}, {Manteiga}, {Su{\'a}rez},
  {Ulla}, {Cunha}, {Metcalfe}, {Kanaan}, {Fraga}, {Costa}, {Giovannini},
  {Fontaine}, {Bergeron}, {O'Brien}, {Sanwal}, {Wood}, {Ahrens}, {Silvestri},
  {Klumpe}, {Kawaler}, {Riddle}, {Reed}, \& {Watson}}]{2003ApJ...594..961M}
{Mukadam}, A.~S., {Kepler}, S.~O., {Winget}, D.~E., {et~al.} 2003, \apj, 594,
  961

\bibitem[{{Mullally} {et~al.}(2008){Mullally}, {Winget}, {Degennaro},
  {Jeffery}, {Thompson}, {Chandler}, \& {Kepler}}]{2008ApJ...676..573M}
{Mullally}, F., {Winget}, D.~E., {Degennaro}, S., {et~al.} 2008, \apj, 676, 573

\bibitem[{{Osaki}(1975)}]{1975PASJ...27..237O}
{Osaki}, J. 1975, \pasj, 27, 237

\bibitem[{{Rappaport} {et~al.}(2015){Rappaport}, {Nelson}, {Levine},
  {Sanchis-Ojeda}, {Gandolfi}, {Nowak}, {Palle}, \&
  {Prsa}}]{2015ApJ...803...82R}
{Rappaport}, S., {Nelson}, L., {Levine}, A., {et~al.} 2015, \apj, 803, 82

\bibitem[{{Redaelli} {et~al.}(2011){Redaelli}, {Kepler}, {Costa}, {Winget},
  {Handler}, {Castanheira}, {Kanaan}, {Fraga}, {Henrique}, {Giovannini},
  {Provencal}, {Shipman}, {Dalessio}, {Thompson}, {Mullally}, {Brewer},
  {Childers}, {Oksala}, {Rosen}, {Wood}, {Reed}, {Walter}, {Strickland},
  {Chandler}, {Watson}, {Nather}, {Montgomery}, {Bischoff-Kim}, {Hansen},
  {Nitta}, {Kleinman}, {Claver}, {Brown}, {Sullivan}, {Kim}, {Chen}, {Yang},
  {Shih}, {Zhang}, {Jiang}, {Fu}, {Seetha}, {Ashoka}, {Marar}, {Baliyan},
  {Vats}, {Chernyshev}, {Ibbetson}, {Leibowitz}, {Hemar}, {Sergeev}, {Andreev},
  {Janulis}, {Mei{\v s}tas}, {Moskalik}, {Pajdosz}, {Baran}, {Winiarski},
  {Zola}, {Ogloza}, {Siwak}, {Bogn{\'a}r}, {Solheim}, {Sefako}, {Buckley},
  {O'Donoghue}, {Nagel}, {Silvotti}, {Bruni}, {Fremy}, {Vauclair}, {Chevreton},
  {Dolez}, {Pfeiffer}, {Barstow}, {Creevey}, {Kawaler}, \&
  {Clemens}}]{2011MNRAS.415.1220R}
{Redaelli}, M., {Kepler}, S.~O., {Costa}, J.~E.~S., {et~al.} 2011, \mnras, 415,
  1220

\bibitem[{{Scuflaire}(1974)}]{1974A&A....36..107S}
{Scuflaire}, R. 1974, \aap, 36, 107

\bibitem[{{Steinfadt} {et~al.}(2010){Steinfadt}, {Bildsten}, \&
  {Arras}}]{2010ApJ...718..441S}
{Steinfadt}, J.~D.~R., {Bildsten}, L., \& {Arras}, P. 2010, \apj, 718, 441

\bibitem[{{Steinfadt} {et~al.}(2012){Steinfadt}, {Bildsten}, {Kaplan},
  {Fulton}, {Howell}, {Marsh}, {Ofek}, \& {Shporer}}]{2012PASP..124....1S}
{Steinfadt}, J.~D.~R., {Bildsten}, L., {Kaplan}, D.~L., {et~al.} 2012, \pasp,
  124, 1

\bibitem[{{Sullivan} \& {Chote}(2015)}]{2015ASPC..493..199S}
{Sullivan}, D.~J. \& {Chote}, P. 2015, in Astronomical Society of the Pacific
  Conference Series, Vol. 493, 19th European Workshop on White Dwarfs, ed.
  P.~{Dufour}, P.~{Bergeron}, \& G.~{Fontaine}, 199

\bibitem[{{Tassoul} {et~al.}(1990){Tassoul}, {Fontaine}, \&
  {Winget}}]{1990ApJS...72..335T}
{Tassoul}, M., {Fontaine}, G., \& {Winget}, D.~E. 1990, \apjs, 72, 335

\bibitem[{{Unno} {et~al.}(1989){Unno}, {Osaki}, {Ando}, {Saio}, \&
  {Shibahashi}}]{1989nos..book.....U}
{Unno}, W., {Osaki}, Y., {Ando}, H., {Saio}, H., \& {Shibahashi}, H. 1989,
  {Nonradial oscillations of stars}, ed. T.~University~of Tokyo~Press

\bibitem[{{Van Grootel} {et~al.}(2013){Van Grootel}, {Fontaine}, {Brassard}, \&
  {Dupret}}]{2013ApJ...762...57V}
{Van Grootel}, V., {Fontaine}, G., {Brassard}, P., \& {Dupret}, M.-A. 2013,
  \apj, 762, 57

\bibitem[{{van Kerkwijk} {et~al.}(2010){van Kerkwijk}, {Rappaport}, {Breton},
  {Justham}, {Podsiadlowski}, \& {Han}}]{2010ApJ...715...51V}
{van Kerkwijk}, M.~H., {Rappaport}, S.~A., {Breton}, R.~P., {et~al.} 2010,
  \apj, 715, 51

\bibitem[{{Vauclair} {et~al.}(2011){Vauclair}, {Fu}, {Solheim}, {Kim}, {Dolez},
  {Chevreton}, {Chen}, {Wood}, {Silver}, {Bogn{\'a}r}, {Papar{\'o}}, \&
  {C{\'o}rsico}}]{2011A&A...528A...5V}
{Vauclair}, G., {Fu}, J.-N., {Solheim}, J.-E., {et~al.} 2011, \aap, 528, A5

\bibitem[{{Winget} {et~al.}(1983){Winget}, {Hansen}, \& {van
  Horn}}]{Wingetet83}
{Winget}, D.~E., {Hansen}, C.~J., \& {van Horn}, H.~M. 1983, \nat, 303, 781

\bibitem[{{Winget} \& {Kepler}(2008)}]{WK08}
{Winget}, D.~E. \& {Kepler}, S.~O. 2008, \araa, 46, 157

\bibitem[{{Winget} {et~al.}(2004){Winget}, {Sullivan}, {Metcalfe}, {Kawaler},
  \& {Montgomery}}]{2004ApJ...602L.109W}
{Winget}, D.~E., {Sullivan}, D.~J., {Metcalfe}, T.~S., {Kawaler}, S.~D., \&
  {Montgomery}, M.~H. 2004, \apjl, 602, L109

\bibitem[{{Zhang} {et~al.}(2016){Zhang}, {Fu}, {Li}, {Ren}, \&
  {Luo}}]{2016ApJ...821L..32Z}
{Zhang}, X.~B., {Fu}, J.~N., {Li}, Y., {Ren}, A.~B., \& {Luo}, C.~Q. 2016,
  \apjl, 821, L32

\end{thebibliography}

\end{document}